\documentclass[twoside]{IEEEtran}  
\usepackage{epsfig}
\usepackage{graphicx}
\usepackage{amsmath}    
\usepackage{verbatim}   

\def\BibTeX{{\rm B\kern-.05em{\sc i\kern-.025em b}\kern-.08em
    T\kern-.1667em\lower.7ex\hbox{E}\kern-.125emX}}

\begin{document}

\title{Performance of the Caltech Submillimeter \\
Observatory Dual-Color 180-720~GHz Balanced \\ SIS Receivers}

\author{J. W. Kooi, R. A. Chamberlin, R. Monje, A. Kov\'{a}cs, F. Rice, H. Yoshida, B. Force, K. Cooper,  \\
D. Miller, M. Gould, D. Lis, B. Bumble, R. LeDuc, J. A. Stern, and T. G. Phillips.
\thanks{This work is supported in part by NSF grant $^\#$ AST-0838261.}
\thanks{ J. W. Kooi, R. A. Chamberlin, R. Monje, A. Kov\'{a}cs , F. Rice, H. Yoshida, K. Cooper, B. Force, 
D. Miller, D. Lis, and T. G. Phillips are with the Submillimeter Astronomy and Instrumentation Group, 
California Institute of Technology, Pasadena, CA 91125, USA. e-mail: kooi@caltech.edu.}
\thanks{B. Bumble, R. LeDuc, and J. A. Stern are with the Jet Propulsion Laboratory (JPL), Pasadena, CA 91109.}
\thanks{M. Gould is with Zen Machine \& Scientific Instruments, Lyons, CO 80540.}
}


\markboth{IEEE Transactions on Terahertz Science and Technology}
{KOOI {\em \lowercase{et al.}}: Caltech Submillimeter Observatory Dual-Color  Balanced SIS Receivers }

\maketitle

\begin{abstract}
We report on balanced SIS receivers covering the astronomical important 180$-$720 GHz submillimeter
atmospheric window. To facilitate remote observations and automated spectral line surveys,
fully synthesized local oscillators are employed. High-current-density Nb-AlN-Nb superconducting-insulating-superconducting
(SIS) tunnel junctions are used as the mixing element. 
The measured double-sideband (DSB) 230 GHz receiver noise temperature, uncorrected for optics loss, ranges from 
50~K at 185~GHz, 33~K at 246~GHz, to 51~K at 280~GHz.  In this frequency range the mixer has a DSB conversion gain 
of 0~$\pm$~1.5~dB. The measured 460~GHz double-sideband receiver noise temperature, uncorrected for optics loss,
is 32~K at 400~GHz, 34~K at 460~GHz, and 61~K at 520~GHz. Similar to the 230 GHz balanced mixer, the DSB mixer 
conversion gain is 1~$\pm$~1~dB.  To help optimize performance, the mixer IF circuits and bias injection 
are entirely planar by design. 
Dual-frequency observation, by means of separating the incoming circular polarized electric field into
two orthogonal components, is another important mode of operation offered by the new facility instrumentation. 
Instrumental stability is excellent supporting the LO noise cancellation properties of the balanced mixer configuration. 
In the spring of 2012 the dual-frequency 230/460 SIS receiver was successfully installed at Caltech Submillimeter Observatory (CSO), Mauna Kea, HI.
\end{abstract}

\vspace{0.2cm}

\section{Introduction} \label{sect:intro}  
\IEEEPARstart{T}he Caltech Submillimeter Observatory (CSO) is located on top of Mauna Kea, Hawaii, at an altitude of 4.2~km. 
To facilitate deep integrations, stable baselines, and automated high 
resolution spectral line surveys significant efforts \cite{kooi_2012, kooi_2007} have been expanded to 
develop a family of synthesized, remote programmable dual-color balanced SIS receivers. 
To this effect four tunerless 
balanced-input waveguide receivers have been developed to cover the important 180$-$720~GHz frequency range \cite{spie2002, spie2004}.
The new heterodyne facility instrumentation allows dual-frequency (two-color) observations in the 230/460~GHz and 345/660~GHz atmospheric 
windows. The 230/460 balanced receiver system has been installed 
and is operational at the observatory since May 2012. 
Unfortunately, deployment of the 345/650 balanced receiver(s) is presently on hold due to funding difficulties.

Dual-frequency observation is an important mode of operation offered by the new facility instrumentation. Dual band observations
are accomplished by separating the horizontal (H) and vertical (V) polarizations of the incoming signal and routing 
them via folded optics to the appropriate polarization sensitive balanced mixer. 
Scientifically this observation mode facilitates pointing for the higher receiver band under mediocre weather conditions and a doubling of scientific throughput under good weather conditions. 

Balanced configurations were chosen for their inherent local oscillator (LO) spurious tone and
amplitude (AM) noise cancellation properties. (It was also judged 
to be an optimal compromise between scientific merit and finite funding). Unique to the CSO, wide RF bandwidth is favored \cite{kooi_2007}, 
allowing the same science to be done with fewer instruments. In all the upgrade covers ALMA bands~5$-$9. 

In principle, the balanced receiver configuration has the advantage that common mode amplitude noise in the local oscillator system is canceled, 
while at the same time utilizing all available LO power.  Both of these features afford automation over unprecedented wide
RF bandwidth, covering the entire 180$-$720~GHz submillimeter atmospheric windows with just four LOs. 
   
Receiver noise temperatures and in situ measured instrumental Allan Variance stability times are excellent and are consistent
with the use of balanced receiver technology.

The SIS junctions are capable of a 13~GHz bandwidth, though due to band limiting isolators and low noise amplifiers the operational
IF bandwidth of the CSO receivers is presently 4$-$8~GHz. 

To maximize the RF bandwidth, we explore the use of high-current-density AlN-barrier SIS technology combined with 
a broad bandwidth full-height waveguide to thin-film microstrip transition \cite{kooi_probe}.
Compared to AlO$_x$-barriers, advantages of AlN tunnel barriers
are a low $\omega$RC product (increased RF bandwidth) and enhanced chemical robustness.
Even if optimal RF bandwidth is not a requirement, a low $\omega$RC product provides a more homogeneous 
frequency response and increased tolerance to errors in device fabrication. 

To process the required IF bandwidth, the CSO has acquired a Fast Fourier Transformer Spectrometer (FFTS) 
from Omnisys Instruments, Sweden. \footnote{
Omnisys Instruments AB., August Barks gata 6B, SE-421 32 V\"{a}stra Fr\"{o}lunda SWEDEN. [Online] Available: http://www.omnisys.se}
This spectrometer facilitates 8 GHz of processing bandwidth with a resolution of 268 KHz/channel, or 3724 channels/GHz. 
The 8 GHz Omnisys FFTS comes in a 19" rack and has two built-in IF processor modules (4$-$8 GHz each), an 
embedded controller module, a synchronization module, and power supply. 

\section{The single-balanced mixer}
\label{sect:Balmixers}

\begin{figure}[t!]
\centerline{
\includegraphics[width=90mm]{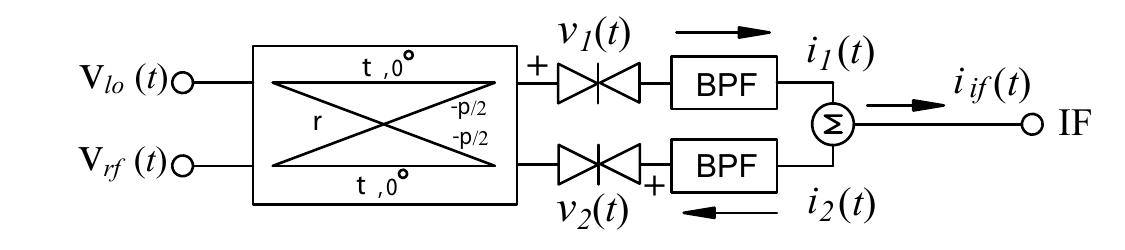}
}
\caption{LO and RF currents in an antipodal biased (single) balanced mixer. In practice, the summing node in the IF 
can be implemented with an in-phase power combiner \cite{Wilkinson} or 180$^\circ$ IF hybrid.
In the case of the CSO mixers the RF input hybrid is 90$^\circ$ with all IF circuitry planar by design. This 
facilitates optimal control of both amplitude and phase. The band pass filter (BPF) is 3$-$9~GHz. 
For further detail we refer the reader to \cite{kooi_2012}.
}
\label{Balmixer}
\end{figure}

As discussed in \cite{kooi_2012}, the single balanced mixer can be formed by connecting 
antipodal biased (SIS) mixers to a 180$^\circ$ or 90$^\circ$ input hybrid. 
The 180$^\circ$ hybrid, though having
superior fundamental and intermodulation product suppression capabilities, is larger and more difficult to implement
at submillimeter frequencies. At submillimeter and terahertz frequencies the harmonic and 
intermodulation products are however severely attenuated by the inherent device capacitance of the mixing element. 
For this reason, submillimeter or terahertz mixers may be configured with quadrature hybrids (Fig. \ref{Balmixer})
\cite{Serizawa2008, Boon-Kok2012} rather than the larger and more complex 180$^\circ$ hybrids couplers.

\noindent
In \cite{kooi_2012} the amplitude rejection of a balanced mixer relative to an ideal single-ended mixer was derived as 

\begin{equation}  
\label{NRdB}
NR(dB) = -20 \cdot log\Bigl[1 - \sqrt{G_m} G_{h} \cos(\Delta\varphi)\Bigr].
\end{equation}

\begin{figure}[t!]
\centerline{
\includegraphics[width=80mm]{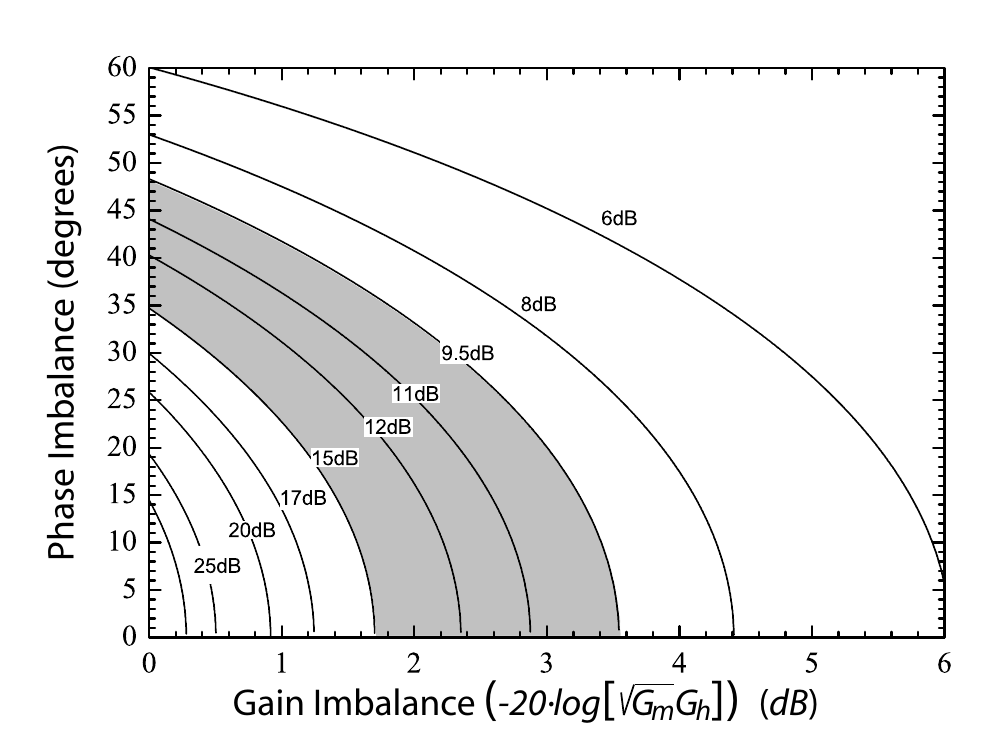}
}
\caption{Amplitude rejection of a balanced mixer relative to an ideal single-ended mixer. The 230~GHz and 460~GHz balanced receivers 
achieve spurious noise amplitude rejection ratio's ranging from 9.5~dB to 15~dB. See also Figs.~\ref{230460Spurs} and \ref{NR}.
}
\label{am-rejection}
\end{figure}

\noindent
which is graphically depicted in Fig. \ref{am-rejection}. Here $\sqrt{G_m}$ is the mixer gain imbalance, $G_h$ the quadrature hybrid imbalance, and 
$cos(\Delta\varphi)$ is the combined phase error of the RF hybrid, device placement, wire bond length, and IF summing node. As will be discussed
in section \ref{spur-rejection}, the measured spurious rejection of the 230~GHz \& 460~GHz balanced mixers is 9.5$-$15~dB.

\section{Hardware}
\subsection{Instrument Configuration}
\label{sect:Instrument-Configuration}

\begin{figure*}[t!]
\centerline{
\includegraphics[width=105mm]{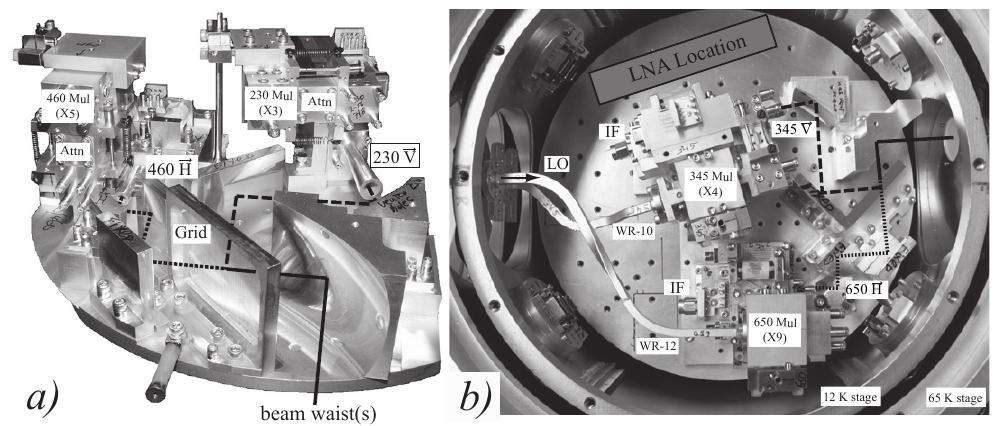}
}
\caption {{\it a)} 230/460~GHz focal plane unit (FPU) with associated balanced mixers, multiplier hardware, and optics. 
{\it b)} 345/650~GHz focal plane unit mounted in the cryostat. Due to the very confined 180~mm diameter cold work surface, 
folded optics is used to re-focus the mixer corrugated feedhorn waist(s) to that of the telescope. The LO signal is coupled 
to the multipliers via WR-10 and WR-12 waveguides from the rear.
}
\label{incryostat}
\end{figure*}

Fig \ref{incryostat} shows two views of the CSO dual-frequency receiver configurations. One cryostat houses 
the 180$-$280~GHz~/~400$-$520~GHz balanced mixers while a second cryostat houses the 
280$-$420~GHz~/~580$-$720~GHz focal plane unit (PFU).

The 63$-$105~GHz LO carrier signal enters the cryostat via an 
(inner wall) Au-plated stainless steel waveguide (WR-10/12). The submillimeter multipliers \footnote{
Virginia Diodes Inc., 979 2nd Street SE, Suite 309  Charlottesville, VA 22902, USA. [Online] Available: http://http://vadiodes.com//}
are mounted on the 15~K stage of a Precision Cryogenics \footnote{
Precision Cryogenics Systems Inc., 7804 Rockville Road, Indianapolis, IN 46214, USA. [Online] Available: http://www.precisioncryo.com/}
hybrid cryostat. The latter is important in that multipliers are inefficient with most of the RF input power converted into heat. 

Each cryostat receives two (orthogonally polarized) beams from the sky,
which are routed via a cold wire-grid to the appropriate mixer. Keeping the RF path length to a minimum the calculated intrinsic mixer block
waveguide loss (4K) ranges from 0.18~dB (24.2~mm)  at 230~GHz to 0.34~dB  (9.3~mm) at 650~GHz.
This technique facilitates dual-frequency (2 color) observations, improves observing efficiency, and assists 
pointing of the high frequency receivers in mediocre weather.

We estimate that the required LO pump power ranges from 100-1000~nW for each (twin) SIS junction 
depending on LO frequency ($\alpha$=$eV_{lo}/h\nu$~$\sim$~0.78 on average) \cite{Tucker-Feldman}. And 
since two SIS junctions are used as part of the balanced configuration we require, including waveguide loss in the mixer block, $\sim$0.5$-$2.5~$\mu$W of local oscillator power at the mixer LO input port.

Given that the cooled frequency multipliers are 1) able to produce ample LO power over the described frequency bands and 2) increase 
25$-$40\% in efficiency upon cooling, it is necessary 
to add attenuation in the LO-mixer path \cite{kooi_2012}. In practice, this may be accomplished with a directional coupler or fixed tuned (preset)
attenuator. A preset attenuator has the advantage of being simple, relatively inexpensive, and manually adjustable at room temperature. 
The effect of employing a cooled attenuator is similar to the use of a beam splitter with quasi-optical LO injection; 
it reduces the multiplier-mixer cavity standing wave, and minimizes additive thermal noise from the 
local oscillator. Additional reduction in LO amplitude and spurious noise
is provided by the ``noise canceling properties" of the balanced mixer as observed from Eq.~\ref{NRdB}.

\subsection{Waveguide to Thinfilm Microstrip Transition}

Traditionally the majority of SIS waveguide mixers employ planar probes that extend all the way
across the waveguide \cite{Woody85}-\cite{Walker}. Unfortunately, the ``double-sided" 
(balanced) probe exhibits a rather poor RF bandwidth ($\leq$~15\%), when constructed in full-height waveguide.
When the height of the waveguide is reduced by 50\%,
the probe's fractional bandwidth improves dramatically to a maximum of about 33\%.
These results can be understood in that the double-sided probe is essentially a planar variation of the well
known Eisenhart and Khan waveguide probe \cite{Eisenhart}. Borrowing from Withington and Yassin's assessment
\cite{Withington1}, the real part of the probe's input impedance is influenced in a complex way
by the parallel sum of individual non-propagating modal impedances, and as such, is frequency dependent.
By lowering the height of the waveguide, the effect of non-propagating modes may be reduced \cite{Tong}-\cite{Jackson}.

\renewcommand{\arraystretch}{0.8}
\begin{table} [t!]
\begin{center}
\caption{Waveguide Transition Parameters of Fig.~\ref{junctionlayout}}
{\begin{tabular}{l c c c c }
\hline
\hline
\rule{-0.7ex}{1.7ex} 
\bf{Parameter/band (GHz)} & 230 & 345 & 460 & 650\\
\hline
\rule{-0.7ex}{1.7ex}  
Quartz Substrate Thickness \emph{($\mu$m)} & 50 &  50 & 50 &  50 \\
Waveguide a-dimension \emph{($\mu$m)} & 889 &  579 & 450 &  310 \\
Waveguide b-dimension \emph{($\mu$m)} & 414 &  290 & 211 &  145\\
Probe radius \emph{($\mu$m)}  & 170 &  112 & 86 &   59 \\
Substrate width \emph{($\mu$m)} & 304 &  204 & 152 &   111 \\
Height above Substrate \emph{($\mu$m)} & 38 &  25 & 16 &  16 \\
Height below Substrate \emph{($\mu$m)}  & 100 &  76 & 50 &  41 \\
Backshort-substrate \emph{($\mu$m)}   & 205 &  102 & 70 &   51 \\
Backshort radius \emph{($\mu$m)} & 100 &  71 & 50 &   50 \\
Probe impedance locus \emph{($\Omega$)} & 49-$i$0 &  48+$i$4 & 46+$i$3 &   42+$i$2 \\
\hline
\hline
\end{tabular}
}
\end{center}
\label{probe}
\end{table}

An alternative approach is to use an asymmetric probe that does not extend all the way across the waveguide. 
For this kind of probe, the modal impedances add in series. The real part of the input impedance
depends only on the single propagating mode and is relatively frequency independent. 
These probes are typically implemented in full-height waveguide, which minimizes conduction loss and reduces
the complexity of fabrication. A rectangular version of the ``one-sided" probe has been used quite extensively by 
microwave engineers \cite{Weinreb, Heuven}, was introduced to the submillimeter community by Kerr {\it et al.}
\cite{Kerr} in 1990, and is currently part of the baseline design for ALMA band 3 and 6 \cite{Alma381, Pan} amongst others.
The radial probe waveguide to thinfilm microstrip transition employed here
represents an attempt to extend the use of radial modes to the
waveguide coupling problem \cite{Withington2}. Design parameters and radial probe dimensions 
for all four waveguide bands are provided in Table I.  Simulations indicate that misalignment errors are
to be kept less than 3$-$4\% of the waveguide height as misalignment of the probe varies 
the 'effective' radius of the probe, thereby altering the shape (bandwidth) of the probe's response \cite{kooi_probe}.

\subsection{High current density AlN barrier SIS junctions}
\label{sect:AlN-SIS}

To facilitate the CSO heterodyne upgrade a suite of high-current-density  AlN-barrier 
niobium SIS junctions (4 bands) have been fabricated by JPL \cite{spie2004}. 
These devices have the advantage of increasing the mixer instantaneous RF bandwidth 
while minimizing absorption loss in the mixer normal or superconducting 
thinfilm front-end RF matching network.

\renewcommand{\tabcolsep}{0.17cm}
\begin{table} [t!]
\begin{center}
\caption{Twin SIS Junction Parameters}
{\begin{tabular}{l c c c c c}
\hline
\hline
\rule{-0.7ex}{1.7ex} 
& \footnotesize{Measured} & \footnotesize{Measured} & \vline & \footnotesize{Design} & \footnotesize{Design} \\
\bf{Band} & R$_{n}$ (\emph{$\Omega$)} & R$^{2mV}_{sg}$ / R$_n$ & \vline & R$_n$ (\emph{$\Omega$)} & Area ($\mu$m$^2$)  \\
\hline
\rule{-0.7ex}{1.9ex}  
230~\emph{GHz} & 6.2 & 12.75 & \vline & 6.33 $\pm$ 10\% & 0.6 \\
345~\emph{GHz} & 5.1 & 10.86 & \vline & 5.43 $\pm$ 10\% & 0.7 \\
460~\emph{GHz} & 5.2 & 12.54 & \vline & 5.43 $\pm$ 10\% & 0.7 \\
650~\emph{GHz} & 4.1 & 12.39 & \vline & 4.22 $\pm$ 10\% & 0.9 \\
\hline
\hline
\end{tabular}
}
\end{center}
\label{junctionpar}
\end{table}

The tunnel junctions under discussion are from batch B030926 and have a measured R$_n$A product 
of 7.6 $\Omega\mu$m$^2$. At the CSO, on top of Mauna Kea, a lower LHe bath temperature (3.67~K) 
results in a subgap leakage current reduction of $\sim$10\%. Relevant twin-SIS junction parameters
are shown in Table II. For a description of the device fabrication we refer to \cite{kooi_2007}. 

\begin{figure}[t!]
\centerline{
\includegraphics[width=90mm]{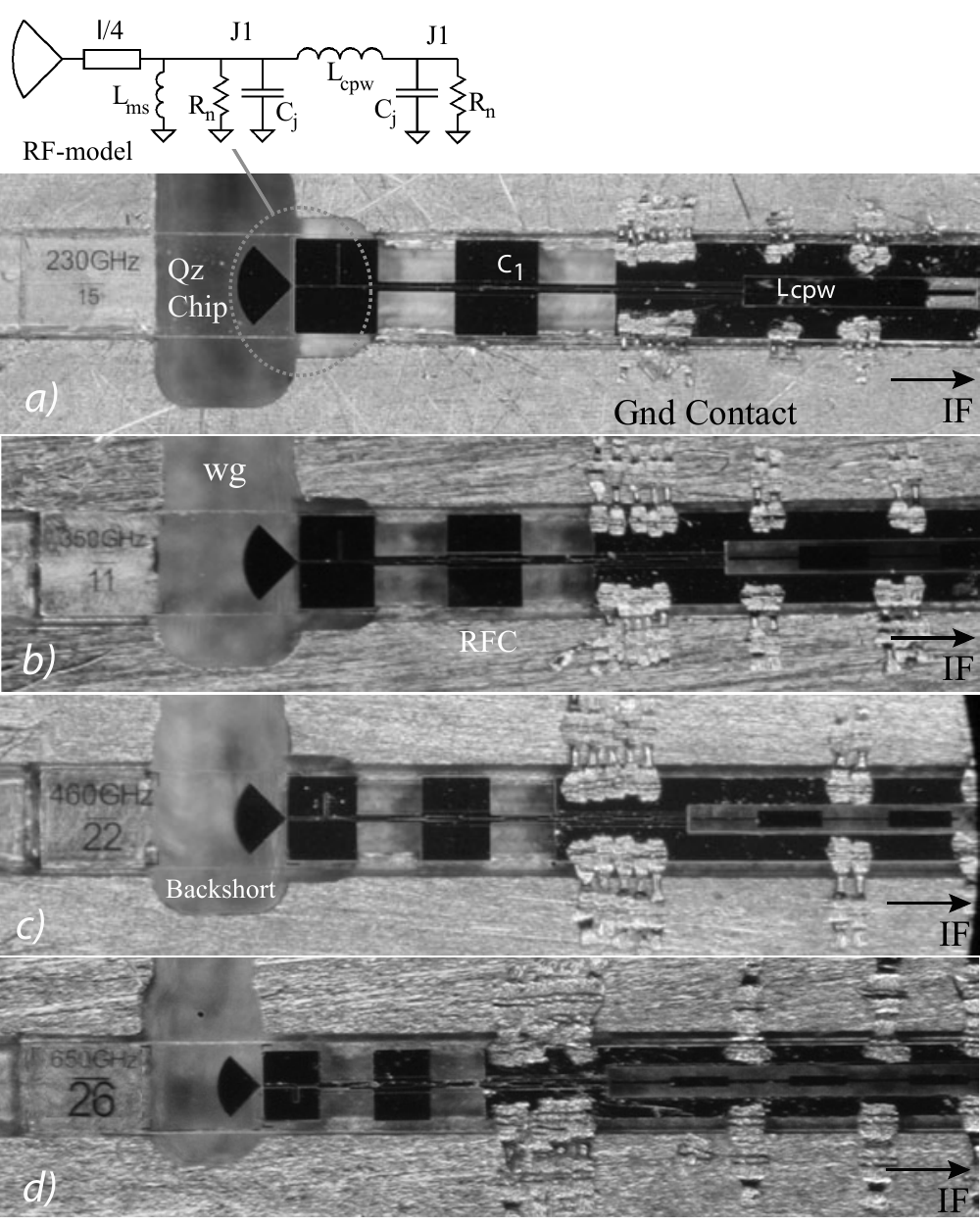}
}
\caption{Chip layout and model of the twin-junction RF tuning  circuit (top). \protect\\
{\it a)} 230~GHz, {\it b)} 345~GHz, {\it c)} 460~GHz, {\it d)} 650~GHz. The radial probe waveguide to thinfilm microstrip transition 
is visible on the left side.} 
\label{junctionlayout}
\end{figure}

\renewcommand{\arraystretch}{1.1}
\begin{table} [t]
\begin{center}
\caption{Waver Parameters}
{\begin{tabular}{l c c }
\hline
\hline
\bf{Parameter} & Measured & Design\\
\hline
C$_s$ (\emph{fF/$\mu$m$^2$}) & 80 & 80 $\pm$~10\% \\
R$_n$A (\emph{$\Omega\mu$m$^2$}) & 7.63  & 7.6 $\pm$~15\% \\
J$_c$ (\emph{kA/cm$^2$}) & 24.9 & 25 $\pm$~15\% \\
V$_{gap}$ (\emph{mV}) & 2.69-2.79  & 2.8 $\pm$~5\% \\
$\delta$V$_{gap}$ (\emph{$\mu$V}) & 50-70  & ---- \\
Nb Top (\emph{nm}) & 420  & 400 $\pm$~20\%  \\
SiO (\emph{nm)} & 320  & 300 $\pm$~15\%  \\
Nb Bottom (\emph{nm}) & 210  & 200 $\pm$~20\%  \\
\hline
\hline
\end{tabular}
}
\end{center}
\label{waverpar}
\end{table}

\subsubsection{Integrated RF matching}
\label{sect:IntegratedRFmatching}

The new SIS tunnel junctions of Fig. \ref{junctionlayout} all share the same 50~$\mu$m thick quartz wafer \cite{Bumble}. 
This has as benefit that the successful wafer run contains all 
the mixer chips needed for the 180$-$720~GHz facility receiver upgrade. 
Supermix \cite{Supermix}, a flexible software library for high-frequency superconducting circuit simulation, was
used in the design process. 

In general, the
junction characteristics are well matched, with slight variations in the definition of the energy gap and device area.
The depicted devices were selected on merit of matching I/V curves, e.g. normal state resistance ($R_n$), leakage current at 2~mV bias, 
gap voltage (V$_{gap}$=2$\Delta/$e), and sharpness of the energy gap.

A limitation of the quadrature hybrid (balanced) design is that the LO power as a function of frequency is
not necessarily equally split between the two (twin) SIS junctions. From detailed analysis we conclude
that gain imbalance due to device characteristics and
LO power imbalance does not significantly affect the overall balanced mixer performance. 
This is important since it means that the individual SIS junctions may be biased at similar, but opposite polarity. 
The simulation results are derived from harmonic balanced superconducting SIS mixer simulations \cite{Supermix} in combination with 
extensive Sonnet \footnote{SONNET, Sonnet Software Inc. [Online] Available: http://www.sonnetsoftware.com/}
 and HFSS \footnote{HFSS, Ansys Inc. [Online] Available: http://www.ansys.com/}
analysis of the RF and IF mixer circuitry, and have been confirmed by measurement (section \ref{results}).

Based on extensive computer simulations, the twin-junction RF matching network was found to exhibit a
slightly larger RF bandwidth than the more common single-junction RF matching 
network \cite{Maier230,Kooi230492}. The AlN-barrier SIS junction [R$_n$C]$^{-1}$ product is 262~GHz,
significantly larger than the bandwidth afforded by the thin-film waveguide transition. This enables uniform 
conversion gain (with margin) across the band of interest. As part of the AlN-barrier characterization process at JPL, 
the specific junction capacitance was estimated 80~fF/$\mu$m$^2$. 
To minimize saturation ($\delta$V$_{sis} \propto [P_{sig} R_n]^{0.5}$),
while maintaining reasonably sized junction areas, we decided on a 5$-$7~$\Omega$ twin-junction normal state resistance design shunted
by an IF embedded impedance of 14~$\Omega$  (Fig. \ref{IF_performance}). Above 13~GHz the integrated capacitor (C$_2$) 
short circuits the IF signal thereby lowering Z$_{emb}$~$\rightarrow$~0~$\Omega$.
Between dc and $\sim$~1~GHz the IF embedding impedance maybe $>$~14~$\Omega$ however this
is a small fraction of the total available IF bandwidth. Note that the IF embedding impedance is in parallel with 
the LO pumped junction impedance, which from measurements takes on values between 30 to 150 Ohm.  From this, and
assuming a 160~GHz RF noise bandwidth (section \ref{sect:FTSmeasurements}), we calculate a bias voltage variation ($\delta$V$_{sis}$) between
0~K and 300~K loads of 80~$\mu$V rms (or less).  Given this analyses we estimate the gain compression on a hot-load  $\leq$  1\%, which 
is supported by the high Y-factors.

\subsubsection{Integrated IF matching}
\label{sect:IntegratedIFmatching}

\renewcommand{\tabcolsep}{0.2cm}
\begin{table} [t]
\begin{center}
\caption{RF Circuit dimensions of Fig \ref{IF_performance}b}
{\begin{tabular}{l l l l l }
\hline
\hline
\bf{Section/Band} & 230~GHz & 345~GHz & 460~GHz & 650~GHz  \\ 
\hline 
\rule{-0.7ex}{1.7ex}  
S1 (l$\times$w) \emph{$\mu$m} & 130.3$\times$3.0  & 85.8$\times$3.2  & 66.8$\times$3.3  & 45.1$\times$4.0  \\
S2 (l$\times$w) \emph{$\mu$m} & 4.0$\times$5.0  & 4.1$\times$5.0  & 4.15$\times$5.0  &   4.5$\times$5.0  \\
S3 (l$\times$w) \emph{$\mu$m} & 116.6$\times$5.0  & 54.9$\times$5.0  & 35.3$\times$5.0  &   14.5$\times$5.0  \\
S4 (l$\times$w) \emph{$\mu$m} & 2.5$\times$5.0  & 2.5$\times$5.0  & 2.5$\times$5.0  &   2.5$\times$5.0  \\
RFC$_h$ (25~$\Omega$) \emph{$\mu$m} & 127$\times$2 & 83$\times$2  & 64$\times$2  &   43$\times$2 \\
RFC$_l$ (10~$\Omega$) \emph{$\mu$m} & 122$\times$6 & 80$\times$6  & 62$\times$6  &   41$\times$6  \\
CPW (Z$_{h}$/Z$_{l}$) \emph{$\Omega$} & 169/77  & 196/84  & 202/95  &   190/134  \\
RFC sections & 10 & 10 & 10 &  10 \\
CPW sections & 3 & 5 & 7 &  9 \\
C2 \emph{(pF)} & 0.883 & 0.843 & 0.820 &  0.798 \\
\hline
\hline
\end{tabular}
}
\end{center}
\label{wg}
\end{table}

\begin{figure*}[t!]
\centerline{
\includegraphics[width=105mm]{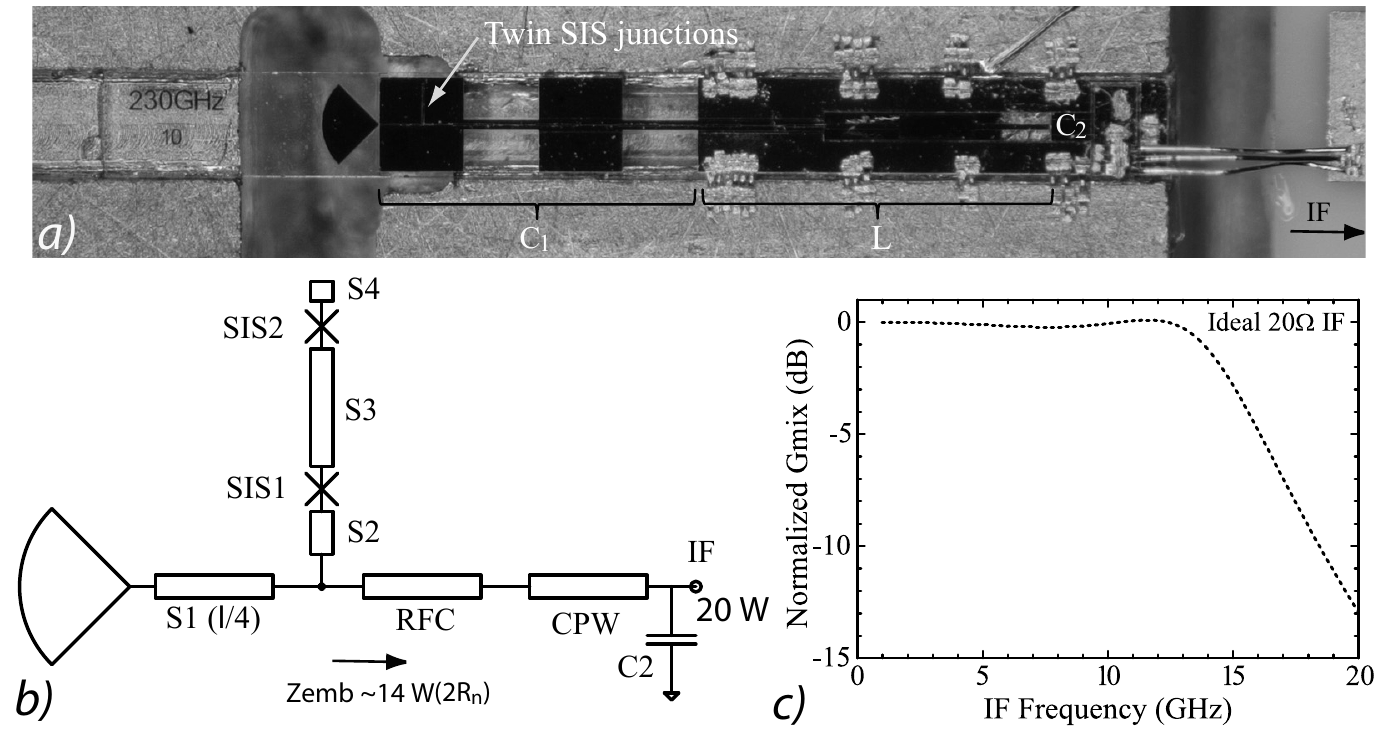}
}
\caption{ 
{\it a)} The IF signal is taken out via a microstrip RF choke (on 300~nm SiO, $\epsilon_r$=5.6) which connects to a 
high impedance CPW transmission line (inductive) and integrated shunt capacitor (C$_2$). {\it b)} This L-C mechanism 
provides a $\pi$ tuning network with the combined capacitance of the probe, twin-junction RF tuning 
structure, and microstrip RF matching network (C$_1$). It also transforms the 20 $\Omega$ IF termination impedance to Z$_{emb}$~$\sim$14~$\Omega$ 
at the junction IF port ($\sim$2R$_n$). {\it c)} The passband is optimized to cover 1$-$13 GHz. To minimize
gain compression, the integrated shunt capacitor also serves to terminate out-of-band broadband noise.
}
\label{IF_performance}
\end{figure*}

Matching to an intermediate IF impedance of 20~$\Omega$ is realized on-chip (Fig. \ref{IF_performance}). The choice of this impedance 
is dictated by the limited available real estate, and the need to minimize gain compression (Z$_{emb}$~$\sim$2R$_n$) \cite{Kerr_sat}.
The mixer design has been optimized for minimum noise temperature and optimal conversion gain, 
while simultaneously regulating the RF and IF input return loss to $\geq$ 8~dB. 
The latter is important as reflections from the RF or IF port can lead to mixer instability.
In Fig. \ref{IF_performance}, we show a photograph of the mixer chip positioned in the waveguide with associated IF model and IF response. 
Short parallel wire bonds provide the ground contact.  To transform the 20~$\Omega$ mixer-chip 
IF output impedance to a 50~$\Omega$ load, an external matching network is employed (Fig. \ref{IF460block}).

\subsection{Wilkinson in-phase Summing Node}
\label{sect:Wilkinson-IF}

\begin{figure}[t!]
\centerline{
\includegraphics[width=90mm]{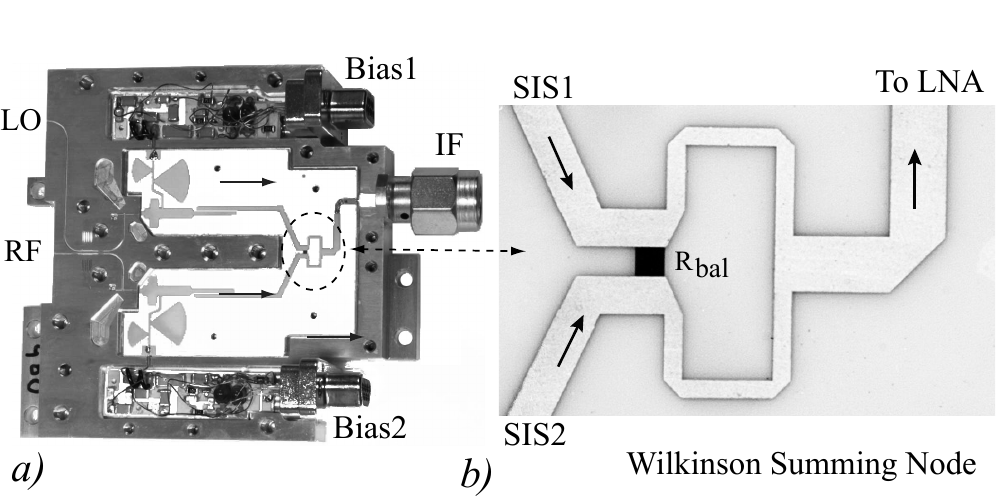}
}
\caption{{\it a)} 460~GHz balanced mixer block with RF quadrature hybrid, IF matching network, dc bias/break,
and Wilkinson summing node. Josephson noise suppression in the SIS tunnel junctions is accomplished by two independent electromagnets
(not shown). {\it b)} Close up of the Wilkinson summing node.
}
\label{IF460block}
\end{figure}

In a mixer configuration, the active device is typically terminated into a desired IF load impedance, the
bias lines EMI-filtered and injected via a bias Tee, and the IF output dc-isolated (Fig.~\ref{IF460block}a).
The balanced mixer has the additional constraint that the individual junction IF output signals need to be combined either 
in phase, or 180$^\circ$ 
out of phase, putting tight limits on the allowed phase error ($<5^\circ$).  Since in our application the SIS junctions
will be biased antipodal (Fig. \ref{Balmixer}) we conveniently combine the bias-Tees, electrical isolation of the IF port, 
band pass filters, IF matching networks, and an in-phase Wilkinson power combiner \cite{Wilkinson} on a single planar circuit.
The 100~$\Omega$ balancing resistor of the Wilkinson power combiner (Fig. \ref{IF460block}b) is a 1\% laser trimmed thinfilm NiCr resistor, 
lithographically deposited on a 635~$\mu$m thick Alumina ($\epsilon_r$=9.8) circuit board  \footnote{American Technical Ceramics, 
One Norden Lane, Huntington Station, NY 11746, USA.}
This compact choice conveniently avoids the use of a physically larger (commercial) 180 degree hybrid.

The IF bandpass filter is comprised of a set of parallel coupled suspended microstrip lines \cite{Menzel}.
For this filter to work, the ground plane directly underneath the filter has been removed, 
and the IF board positioned on top of a machined cutout (resonant cavity). There are several discontinuities 
in this structure. When combined, they form the bandpass filter poles. The advantages are; simplicity of design
(only one lithography step), accurate knowledge of the phase, and reliability. 
The disadvantage is possibly its size, $\lambda_g$/4 ($\sim$6~mm at 6~GHz).

\subsection{IF Noise Characterization}
\label{sect:Andreev}  

\begin{figure*}[t]
\centerline{
\includegraphics[width=105mm]{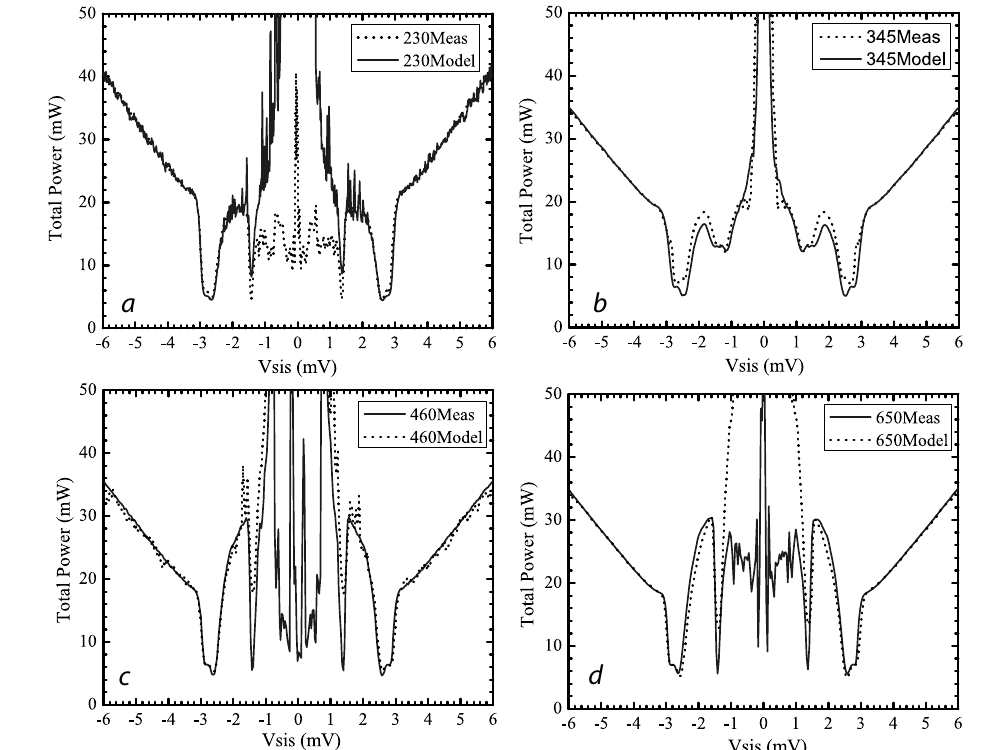}
}
\caption{Measured and modeled IF power for the 230~GHz~(a), 345~GHz~(b), 460~GHz~(c), and 650~GHz~(d) 
balanced mixers. The magnetic field was not necessarily optimized resulting in large deviations due to Josephson oscillations 
near 0~mV. Modeled results are shown in Table V.
}
\label{ShotAll}
\end{figure*}

It is found, not surprisingly, that high-current-density (J$_c$) AlN-barrier tunnel junctions exhibit a 
slightly larger leakage current than commonly used lower-J$_c$ AlO$_x$ tunnel junctions
(R$_{sg}$/R$_n$ ratios of 10$-$14 vs 20$-$35).  
To derive the IF noise contribution of the receiver, and to 
investigate charged quantum transport current by means of multiple Andreev reflections (MAR) 
through pinholes in the AlN barrier we apply a technique described by Dieleman {\it et al.} \cite{Dieleman}. 
In the analysis, the current spectral density below the energy gap ($S_{I}(V)$) is modeled by summing 
the thermalized single-electron tunnel current (I$_{tun}$) with a charged quantum transport current (I$_{mar}$).
In this case

\begin{equation}
\label{Sv}
S_I(V)=2eI_{tun}+2q(V)I_{mar},~~q(V)=(1+2\Delta/eV),
\end{equation}

\noindent
with $2\Delta/e$~=~2.75~mV for our AlN junctions. Rearranging Eq.~\ref{Sv} by defining
r=I$_{tun}$/I with I=(I$_{tun}$+I$_{mar}$) gives

\begin{equation}
\label{Sv2}
S_I(V)=2eI\biggl[1+\frac{2\Delta}{e V}(1-r)\biggr]. 
\end{equation}

\noindent
The noise contribution of a single junction to the IF output is then given by

\begin{equation}
\label{Pif}
P_{IF}=G_{IF}B \biggl[\frac{S_I(V) R_d}{4} (1-\Gamma_{IF}^2)\biggr],
\end{equation}

\noindent
where 

\begin{equation}
\label{gamma}
\Gamma_{IF} = \frac{R_d - Z_o}{R_d + Z_o}
\end{equation}

\noindent
G$_{IF}$ is the IF gain, B the IF bandwidth, R$_d$ the differential resistance obtained from the measured unpumped
I/V curve, and Z$_o$ the IF impedance (20~$\Omega$ in this case). From the analyses we conclude that only about 
1$-$2\% (Table V) of the enhanced subgap noise is due to charged quantum transport (MAR). 

Kerr {\it et. al} has shown \cite{NRAO} that an ideal (0~K) broadband DSB mixer with zero point fluctuations associated
with the signal and image sidebands has a minimum mixer (IF) output noise of half a photon ($h\nu/2k_B$). From
the derived DSB mixer noise temperature (Section \ref{results}) we estimate the (finite temperature) mixer
thermal noise contribution and subgap shot noise, due to leakage current (Table II), to contribute an additional 
half a photon of noise. 

To calculate the IF noise contribution, Rudner {\it et al.} \cite{Rudner}, 
and Woody {\it et al.} \cite{Woody85}, proposed to use the unpumped junction biased above the 
superconducting energy gap as a calibrated shot noise source (Eq. \ref{Tshot}). Studies by
Dubash {\it et al.} \cite{Dubash93, Dubash95} quantitatively verified that the noise current of an unpumped SIS junction above the gap is 
in fact the shot noise associated with the direct current. Hence the current is entirely due to single-electron tunneling
and T$_{shot}$ may be found in the traditional way:

\begin{equation}
\label{Tshot}
T_{shot}=\frac{e R_d I}{2 K_b} coth\biggl(\frac{e V}{2 K_b T}\biggr),
\end{equation}

with $K_b$ Boltzmann constant and

\begin{equation}
S_I(V)=\frac{4 T_{shot} K_b}{R_d} coth\biggl(\frac{e V}{2 K_b T}\biggr)^{-1}=2eI.
\end{equation}

For the balanced mixer and twin-junction design there are 4 SIS junctions. Since the noise contribution of each junction is uncorrelated, 
the resultant mixer output noise is obtained by adding the MAR and shot noise in quadrature (Fig.~\ref{ShotAll}). 
Finally, to compute the IF noise contribution and mixer conversion gain (Section \ref{sect:RxPerformance}) 
we use a technique explained by Wengler and Woody \cite{Wengler87}.

\renewcommand{\arraystretch}{1.1}
\begin{table} [t!]
\begin{center}
\caption{IF Parameters. \protect\\
\scriptsize{$^{\dagger}$Taken at the same gain setting.} 
}
{\begin{tabular}{l c c c }
\hline
\hline
\bf{Band} & T$_{if}$ (K) & MAR (\%) &  Gain (dB) \\
\hline
230~\emph{GHz}$^{\dagger}$  & 3.51 &  0.70 & 67.1 \\
345~\emph{GHz} &  4.62 &  0.20 & 43.1 \\
460~\emph{GHz}$^{\dagger}$  & 3.91 &  2.20 & 67.1 \\
650~\emph{GHz} & 4.74 &  1.50 & 53.0 \\
\hline
\hline
\end{tabular}
}
\end{center}
\label{IFpar}
\end{table}

\section{Receiver Performance}
\label{sect:RxPerformance}

\subsection{Optics} 
\label{sect:optics}
The receiver noise temperature is critically dependent on optical loss 
in front of the mixer. This can be understood from 

\begin{equation}
\label{Treceiver}
T^{DSB}_{rec}=T_{rf} + \frac{T_{mix}}{G_{rf}}+ \frac{T_{IF}}{G_{rf}G^{DSB}_{mix}}.
\end{equation}

\begin{figure*}[t!]
\centerline{
\includegraphics[width=103mm]{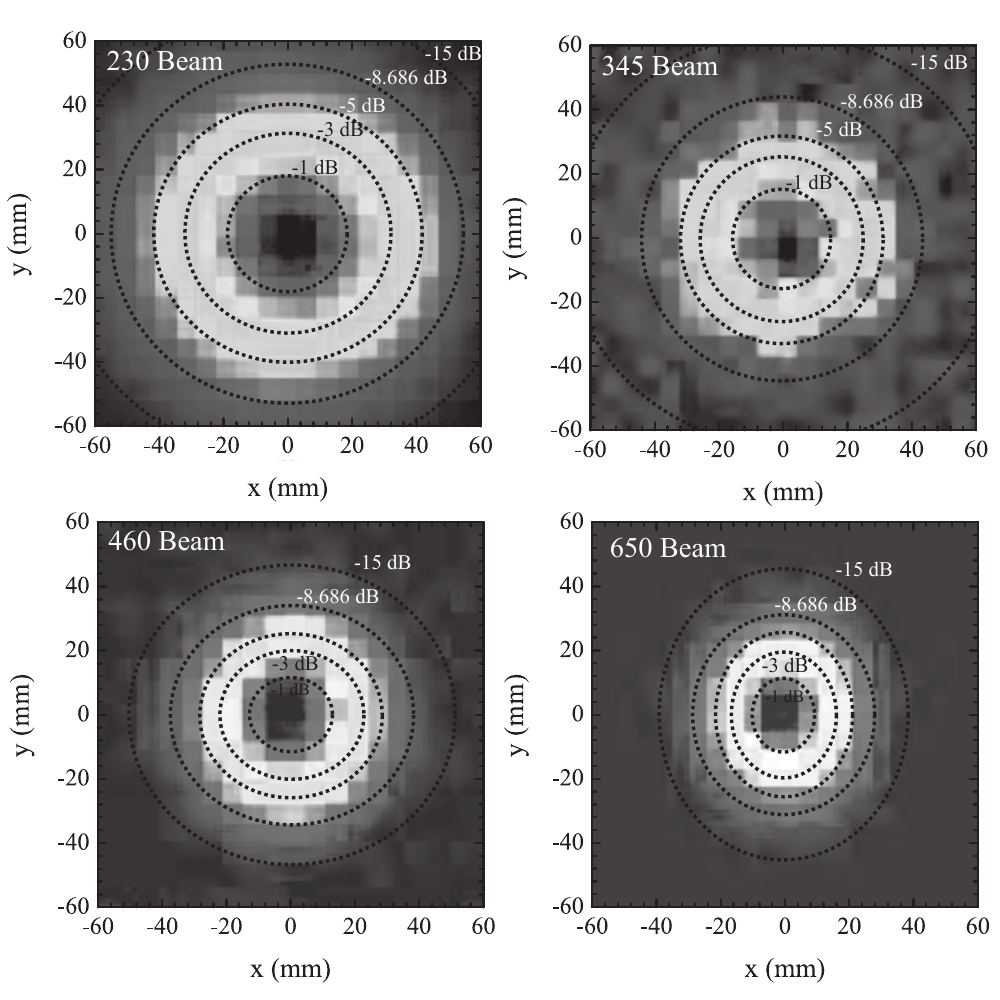}
}
\caption{Beam contours measured in direct detection mode.
The somewhat higher 460~GHz and 650~GHz beam eccentricity is not unexpected given the 
right angle folded optics (Fig.~\ref{incryostat}.)
}
\label{BeamContours}
\end{figure*}

\begin{figure*}[t!]
\centerline{
\includegraphics[width=103mm]{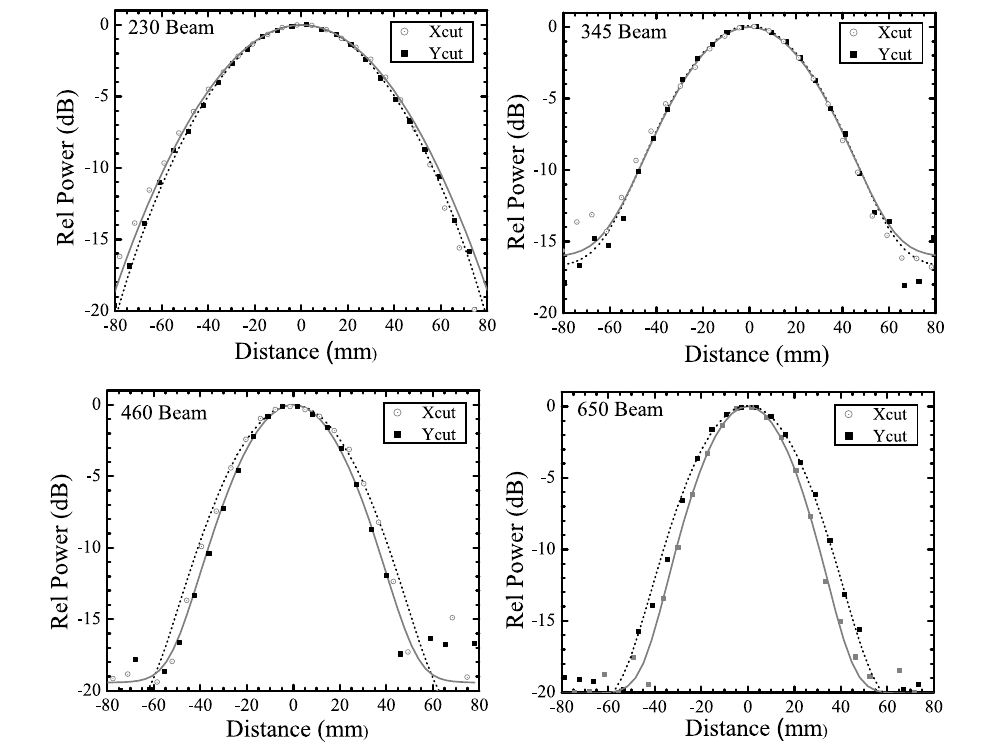}
}
\caption{Cross cuts of the measured 230, 345, 460, 650~GHz beams. Measurements were taken
in direct detection (continuum) mode which limits the SNR to $\sim$18~dB. 
Fit parameters to a single mode Gaussian beam are tabulated in Table VI.
}
\label{GaussianBeams}
\end{figure*}

\noindent
$G^{DSB}_{mix}$ is the double-sideband mixer gain, $G_{rf}$ the front-end optics transmission coefficient,
$T_{rf}$ the optics noise temperature, $T_{IF}$ the IF noise temperature, and $T_{mix}$ the intrinsic mixer noise.
We have minimized the optics noise by careful selection of the vacuum window and infrared blocking filters \cite{kooi_2012}, 
and by use of only cooled reflective optics.

\begin{figure*}[t!]
\centerline{
\includegraphics[width=105mm]{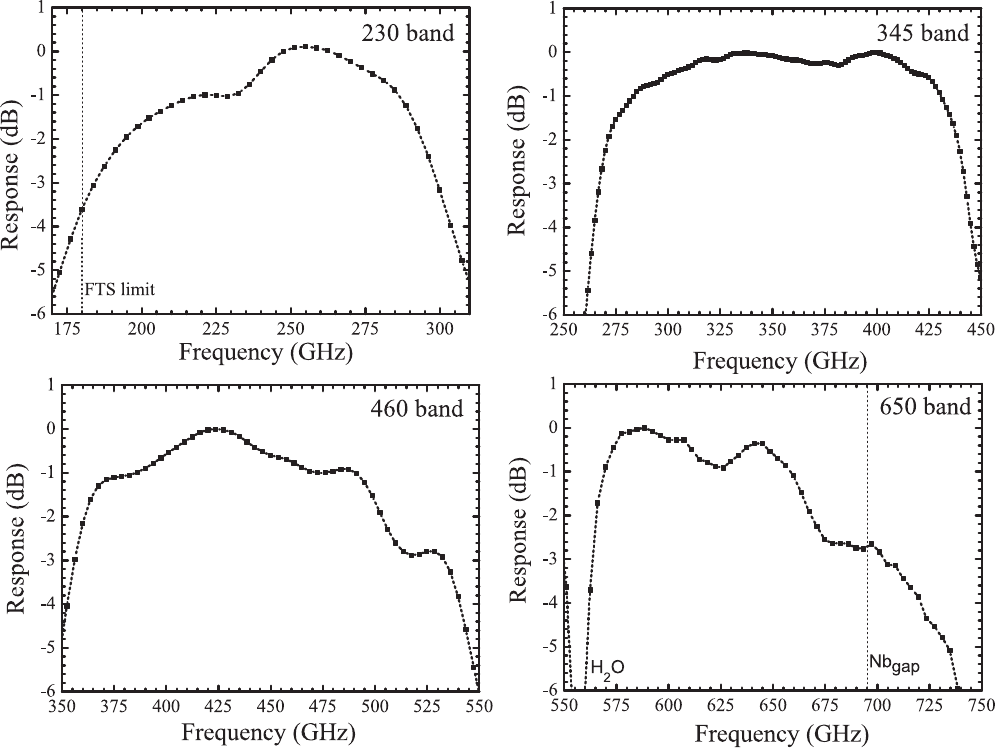}
}
\caption {Normalized direct detection passband response as measured with a Fourier Transform Spectrometer (FTS).
Below ~180~GHz vignetting in the FTS causes the response to roll off.
}
\label{FTSresponse}
\end{figure*}

For optimal RF bandwidth and performance, we use a corrugated feedhorn \footnote{Custom Microwave Inc., 940 Boston Avenue Longmont, CO 80501, USA.
[Online] Available: http://www.custommicrowave.com/} with $\sim$43\% fractional 
bandwidth. The design is based on numerical simulations of a 180$-$280~GHz feedhorn
with 64 sections by J. Lamb \cite{OVRO}. Calculated input return loss of the horn is better than 18~dB, the
cross-polar component less than -32~dB, and the phase front 
error 0.1. The horns have a frequency independent waist \cite{Goldsmith} resulting in a frequency dependent beam
divergence (\emph{f/D=}$\pi w_{o} / 2\lambda$).

In our design the FPU output waist is positioned at the 65~K stage of the cryostat (Fig. \ref{incryostat}).
This allows the use of a 32~mm diameter pressure window (7~$\omega_o$). 

To provide a constant telescope illumination and maximum aperture efficiency, an edgetaper of $\sim$11~dB \cite{Goldsmith} was deemed optimal
given the secondary mirror central blockage.
To achieve this for the CSO Nasmyth focus two focusing elements (elliptical mirrors) are required. In addition, to allow for dual-band 
observations, a wire grid is needed to separate the incoming circular-polarized astronomical signal into the respective 
$H$ and $V$ linear-polarized components. Combining these
requirements with the limited cold surface work space (120~mm) required the use of 'dense' folded optics. The 230~GHz and 345~GHz beams suffer from
less distortion than the 460~GHz and 650~GHz beams as the last focusing mirror could be made of a 37.5$^\circ$ off-axis elliptical mirror
as opposed to a 45$^\circ$ off-axis elliptical mirror. In Figs. \ref{BeamContours}, \ref{GaussianBeams} 
we show the measured direct detection beams. Fit parameters to a single mode Gaussian beam are tabulated in Table VI.

\renewcommand{\arraystretch}{1.0}
\begin{table} [t]
\begin{center}
\caption{FPU Optics Parameters\protect\\
}
{\begin{tabular}{l c c  c c c }
\hline
\hline
\bf{Band} & FWHM \emph{(mm)} & Waist \emph{(mm)} & f \emph{(mm)} & f/D & $e$  \\
\hline
230~\emph{GHz} & 31.644 & 53.82  & 321 & 2.98 & 0.266  \\ [0.05ex]
345~\emph{GHz} & 25.620 & 44.109 & 347 & 3.93 & 0.023  \\ [0.05ex]
460~\emph{GHz} & 21.144 & 36.114 & 312 & 4.32 & 0.461  \\ [0.05ex]
650~\emph{GHz} & 18.264 & 31.172 & 338 & 5.42 & 0.525  \\ [0.05ex]
\hline
\hline
\end{tabular}
}
\end{center}
\label{Multiplication-factors}
\end{table}

\begin{figure}[t!]
\centerline{
\includegraphics[width=90mm]{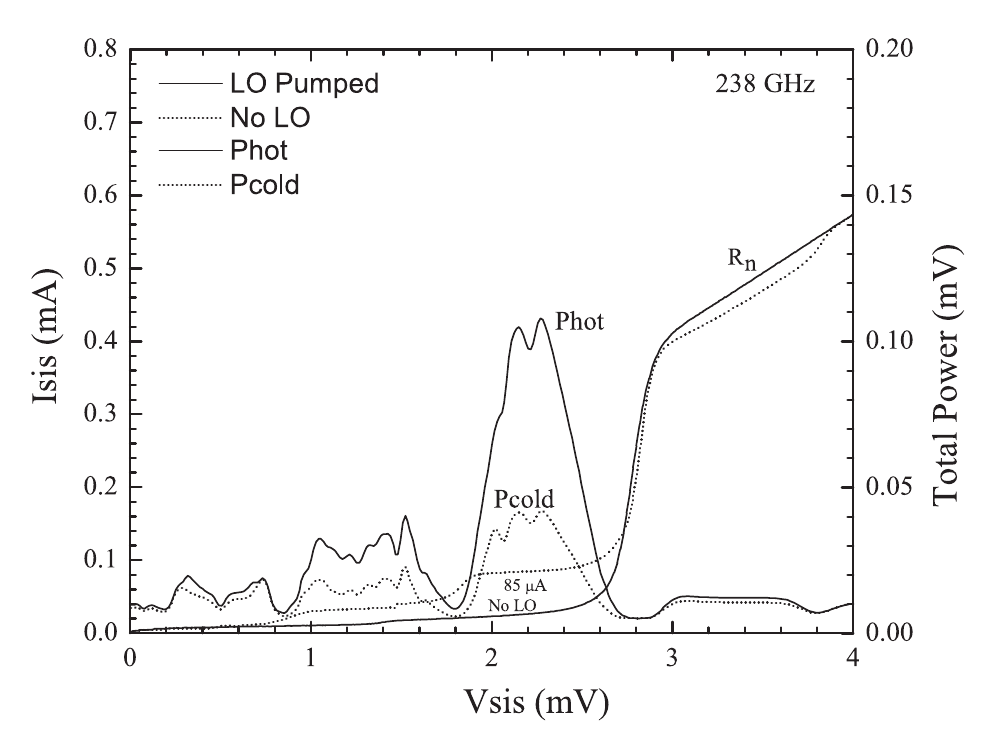}
}
\caption{238~GHz balanced mixer heterodyne response. The
first and second photon steps below the gap are clearly visible. 
Details are provided in Table VII and Fig.~\ref{230Trec}. Optimal bias is 
approximately 2.2~mV.
}
\label{230heterodyne}
\end{figure}

\subsection{Fourier Transform Spectrometer Measurements}
\label{sect:FTSmeasurements}

To investigate the coupling to the twin-SIS junction RF matching network (Fig. \ref{junctionlayout}), 
we have measured the direct-detection response of the mixer with a
Fourier transform spectrometer (FTS). The result is shown in Fig. \ref{FTSresponse}.
Mixer bias is antipodal, approximately {\small 1/2} photon below the energy gap (2$\Delta$/e $\sim$ 2.75~mV).

For the lower frequency bands the RF passband is limited by the corrugated feedhorn and radial probe waveguide transition fractional bandwidth
 ($\sim$ 43\%), whereas for the higher frequency bands the RF passband limit is set by the twin-SIS junction RF matching network  ($\sim$160~GHz).
In addition, the measured direct-detection responses are centered on the respective (designed) passbands. This argues
for the accuracy of the computer simulations \cite{Supermix} and quality of the device fabrication.

\subsection{Heterodyne Results and Discussion} 
\label{results}

In Fig. \ref{230heterodyne} we show the down-converted (heterodyne) 
'hot' and 'cold' load response with associated 
local-oscillator pumped and unpumped I/V curves at $\nu_{LO}$~=~238~GHz. The measured result is 
representative of hundreds of characterization curves taken in automated fashion 
across the 180$-$280~GHz frequency range. For all frequencies
best 230~GHz mixer bias occurs between 2.1$-$2.2~mV (see also Fig. \ref{NR}). 
Optimal LO pump current is 85$-$102~$\mu$A, which is 58$-$75~$\mu$A over the dark current.
From this we calculate that $\alpha_{230} \equiv e V_{LO} / \hbar\omega$ ranges from 0.78$-$0.94,
corresponding to a mean LO pump level of 110~nW.

\begin{figure}[t!]
\centerline{
\includegraphics[width=90mm]{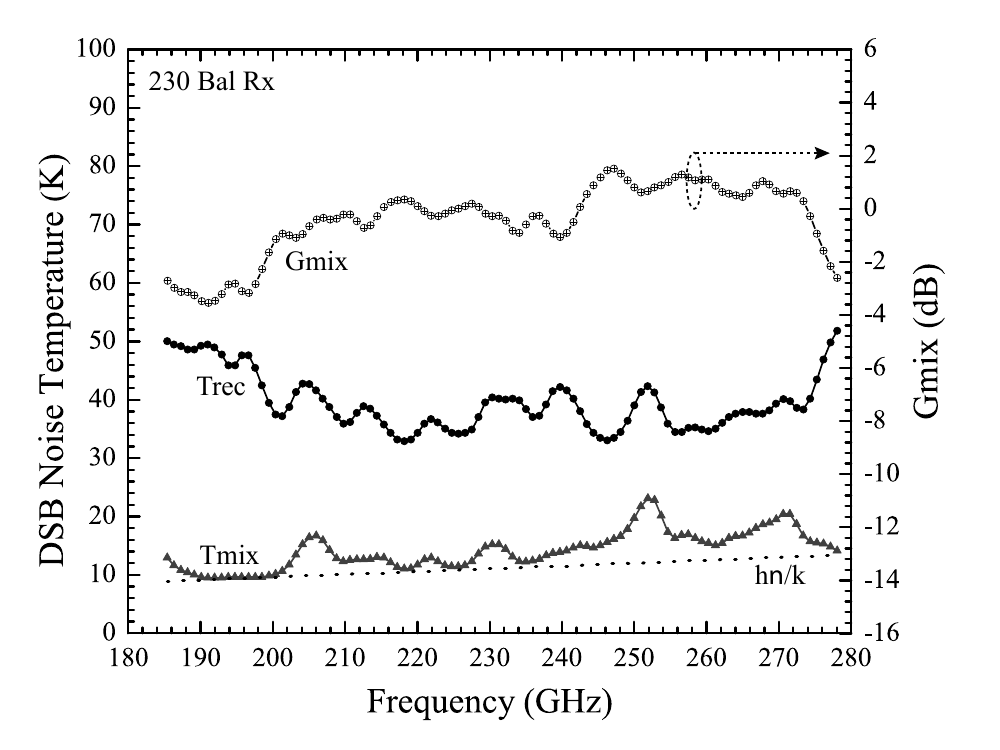}
}
\caption{230~GHz receiver sensitivity and mixer gain as a function of LO frequency. The mixer gain is approximately unity.
Optimal bias occurs between 2.1$-$2.2~mV with T$_{mix} \sim h\nu/k$. The
magnetic field (current) was fixed biased at 7~mA, corresponding to the first Josehpson null.
}
\label{230Trec}
\end{figure}

\begin{figure}[t!]
\centerline{
\includegraphics[width=90mm]{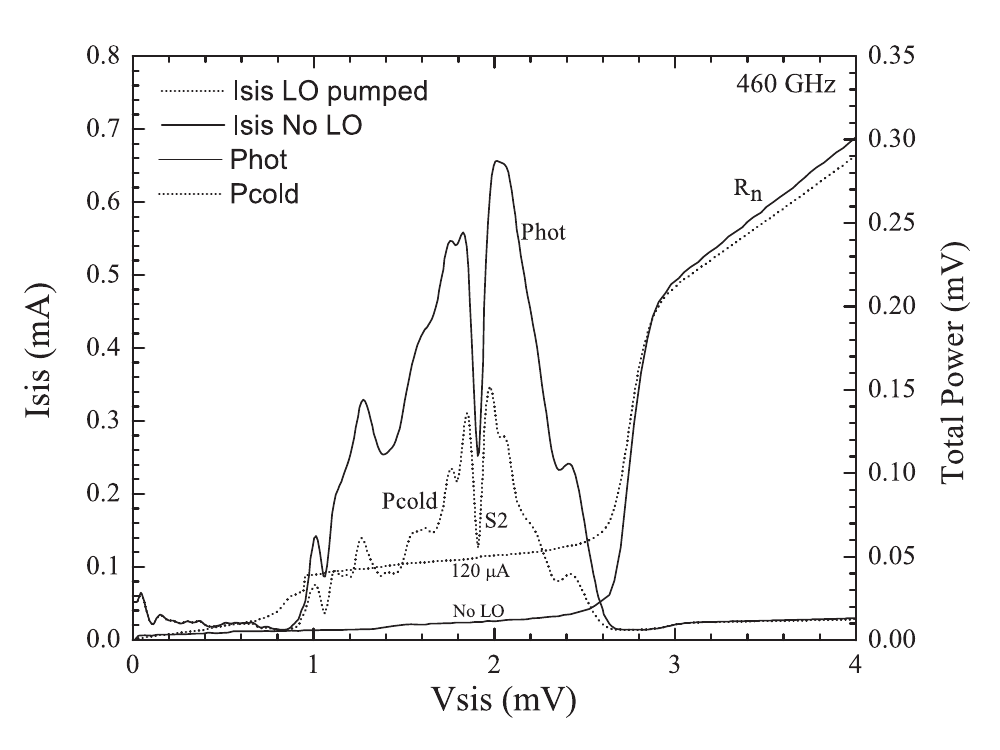}
}
\caption {460~GHz balanced mixer heterodyne response. The 2$^{nd}$
Shapiro step falls typically in the middle of the first photon step below the gap and
is difficult to suppress for the two twin-junctions given a fixed (10~mA) magnetic field setting. 
For bias information see text.
}
\label{460heterodyne}
\end{figure}

For tuning simplicity it was decided to use only one bias setting for both electromagnets (Fig.~\ref{incryostat}b) at all frequencies. 
We note that the superconducting electromagnets were wound (nearly) 'identical' for this reason. In addition, the SIS junction geometry
does not have a 'diamond' like e-beam profile \cite{Karpov} to facilitate homogenous suppression of the ac-Josephson effect (breaking of Cooper pairs).
This could be a future enhancement. Because two twin-SIS junctions are involved it is therefore practically
impossible to perfectly null Josephson oscillations with a common bias setting on both electromagnets (Fig.~\ref{incryostat}b). However
since the location of the Shapiro steps are well known (S$_{n}=nh\nu/2e$, n=1, 2, ...), it is avoided by a computerized bias algorithm.

\begin{table}[t!]
\begin{center}
\caption{Sample of measured 230~GHz receiver parameters.\protect\\
\scriptsize{$^{\dagger}$Includes spillover due to the fast beam (Fig.~\ref{GaussianBeams}).} 
}
{
\begin{tabular}{l l l l }
\hline
\hline
\multicolumn{1}{l}{\bf{Parameters}} & 203~GHz & 238~GHz & 275~GHz \\ 
\hline
\rule{-0.7ex}{2.1ex}  
T$_{rec}^{DSB}$ \emph{(K)} & 41.3 & 40.0 & 41.0\\
$^\dagger$T$_{rf}$ \emph{(K)}  & 21 & 21 & 19\\
T$_{IF}$ \emph{(K)}  & 3.5 & 3.5  & 3.5\\ 
T$_{mix}$ \emph{(K)} & 13.4  & 13.5 & 15.5\\
$G_{mix}^{DSB}$ \emph{(dB)} & -1.09   & -0.56  & -0.95\\ 
$G_{rf}$ \emph{(dB)} & -0.304 & -0.304  & -0.273\\ 
T$_{mix}$/($G_{rf}$) \emph{(K)} & 14.3 & 14.5 & 16.5\\
T$_{IF}$/($G_{rf}G_{mix}^{DSB}$) \emph{(K)} & 4.8 & 4.3 & 4.6\\ 
$\alpha_{230}$  & 0.94 & 0.78 & 0.80\\
\hline
\hline
\end{tabular}
}
\end{center}
\label{results3}
\end{table}

\begin{table}[t]
\begin{center}
\caption{Sample of measured 460~GHz receiver parameters.}
{
\begin{tabular}{l l l l }
\hline
\hline
\multicolumn{1}{l}{\bf{Parameters}} & 406~GHz & 460~GHz & 495~GHz \\ 
\hline
\rule{-0.7ex}{2.1ex}  
T$_{rec}^{DSB}$ \emph{(K)} & 34.0 & 37.1 & 40.2\\
T$_{rf}$ \emph{(K)}  & 8 & 8 & 9\\
T$_{IF}$ \emph{(K)}  & 3.9 & 3.9  & 3.9\\ 
T$_{mix}$ \emph{(K)} & 22.5  & 28.2 & 28.4\\
$G_{mix}^{DSB}$ \emph{(dB)} & -1.03   & 2.38  & 0.76\\ 
$G_{rf}$ \emph{(dB)} & -0.113 & -0.113  & -0.142\\ 
T$_{mix}$/($G_{rf}$) \emph{(K)} & 23.09 & 28.9 & 29.3\\
T$_{IF}$/($G_{rf}G_{mix}^{DSB}$) \emph{(K)} & 5.0 & 2.3 & 3.4\\ 
$\alpha_{460}$ & 0.75 & 0.78 & 0.75\\
\hline
\hline
\end{tabular}
}
\end{center}
\label{results3}
\end{table}

\begin{figure}[t!]
\centerline{
\includegraphics[width=90mm]{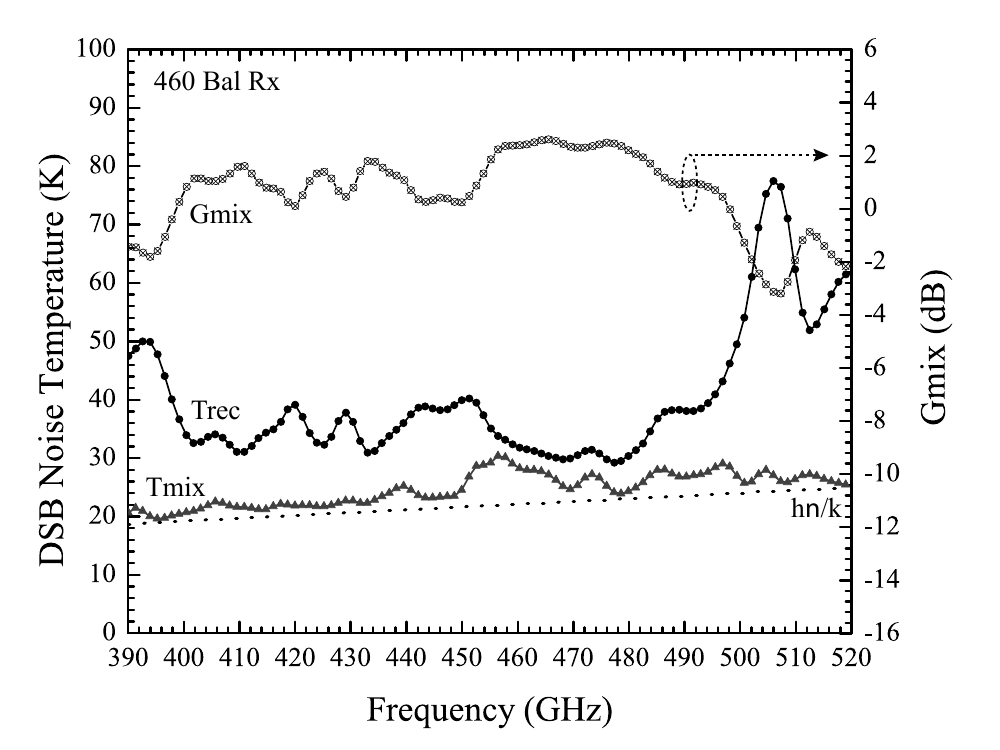}
}
\caption{460~GHz band receiver sensitivity and mixer gain as a function of LO frequency.
The mixer gain is slightly greater then unity and T$_{mix}~\sim$~10\% higher then the quantum noise limit.
}
\label{460Trec}
\end{figure}

\begin{figure*}[t!]
\centerline{
\includegraphics[width=120mm]{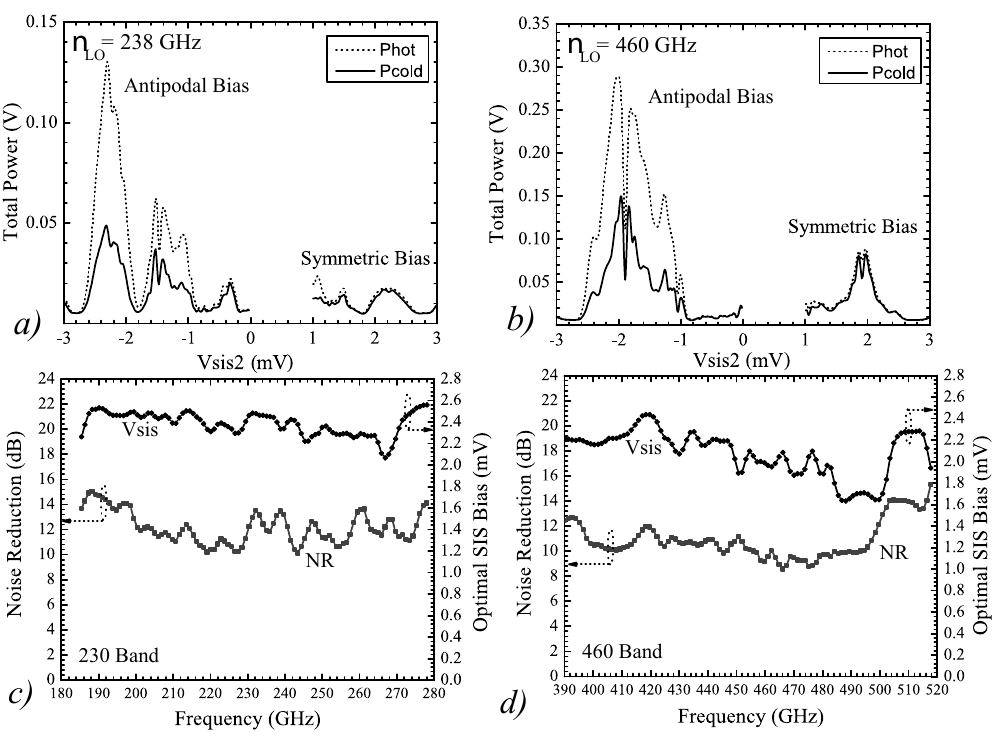}
}
\caption{238~GHz ({\it a}) and 460~GHz ({\it b}) 'hot' and 'cold' load response with the 230~GHz/460~GHz balanced mixer bias swept
either antipodal or symmetric. In all instances SIS junction \#1 was biased positive. {\it c, d)} Derived
amplitude noise reduction and optimal mixer bias (Figs.~\ref{230Trec}, \ref{460Trec}) for both mixers
as a function of frequency. Not surprisingly, the 230~GHz mixer noise reduction
is slightly larger than the 460~GHz mixer noise reduction. Both marginally surpass the estimate of \cite{kooi_2012}.
}
\label{NR}
\end{figure*}

As a general principle the receivers were not biased for maximum mixer conversion gain ($G_{mix}^{DSB}$),
which occurs when the IF output power is optimized. Rather, we developed a global search routine 
and obtained at 100 frequencies across each receiver band the optimal receiver sensitivity 
as a function of LO pumping and SIS bias. The results are shown in Figs.~\ref{230Trec}, \ref{460Trec}.

\begin{figure*}[t!]
\centerline{
\includegraphics[width=105mm]{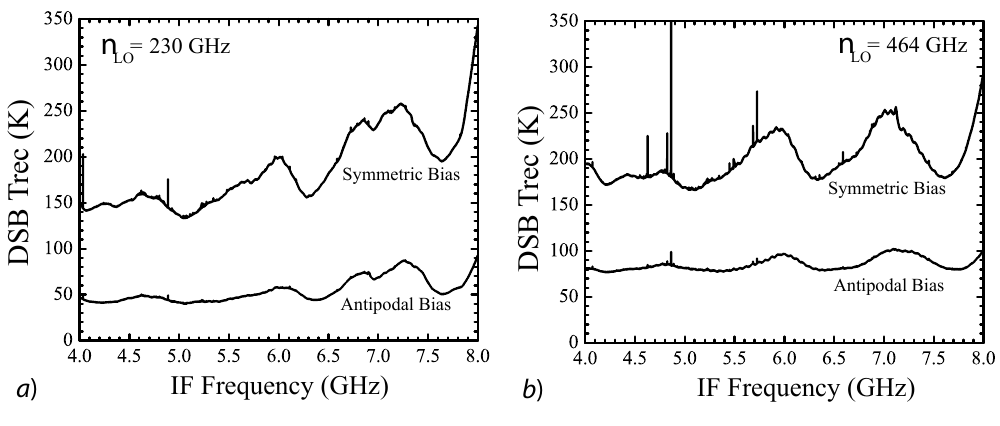}
}
\caption{Spurious rejection with the 230 ({\it a}) and 460 ({\it b}) junctions biased symmetric (top curves) and antipodal (bottom curves).
A second synthesizer was used to inject the spurious tones. See text for detail.
}
\label{230460Spurs}
\end{figure*}

In the measurements, the input load temperatures were defined using 
Callen \& Welton formalism \cite{Callen, NRAO}, where the vacuum zero-point fluctuation noise 
is included in the blackbody radiation temperature. At the frequencies of interest this approaches the
Rayleigh-Jeans limit.

To understand the optics loss in front of the mixer, we employ
a technique, commonly known as the ``intersecting-line technique", described 
by Blundell {\it et al.} \cite{Blundell} and Ke and Feldman \cite{Feldman}. We find between 180$-$280~GHz
a front-end equivalent noise temperature (T$_{rf}$) of 18$-$21~K and between 390$-$520~GHz
a front-end equivalent noise temperature of 8$-$9~K. These losses include vignetting and spillover
and are considerably worse for the 230 receiver then the 460 receiver due to the fast optics. In addition the 
AR coated HDPE pressure window was optimized for the 460~GHz atmospheric band \cite{kooi_2012}. 
A detailed breakdown of the noise budget at three frequencies across the mixer
band is provided in Tables $VII~\&~VIII$.

In Fig. \ref{460heterodyne} we show the measured heterodyne response and associated 
local-oscillator pumped and unpumped I/V curves at $\nu_{LO}$~=~460~GHz. The measured result is again
representative of hundreds of characterization curves taken.
For all frequencies best mixer bias, avoiding Josephson oscillations, falls in two ranges: 
$\pm[1.6-2~mV], \pm[2.2-2.4~mV]$ depending on the frequency of operation (Fig. \ref{460heterodyne}).
Optimal LO pump current is 105$-$113~$\mu$A, which is 76$-$87~$\mu$A over the dark current.
From this we calculate that $\alpha_{460}$~=~0.75$-$0.78 which corresponds to a mean LO pump level of 394~nW.

\subsection{Spurious  Signal and Amplitude Noise Rejection} 
\label{spur-rejection}  

An important motivation of using balanced mixers is the inherent spurious signal and amplitude noise cancellation characteristics
this type of mixer configuration offers. It is of interest therefore to 
establish the actual vs. modeled noise rejection.  It should be noted that a second, and possibly equally important,
motivation for balanced mixers is the efficiency with which the local oscillator carrier signal may be injected \cite{kooi_2012}.

\begin{figure*}[t!]
\centerline{
\includegraphics[width=105mm]{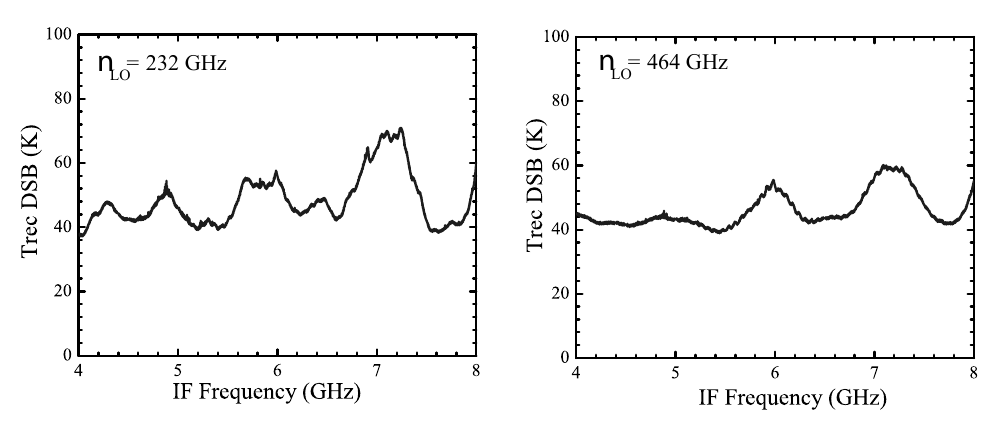}
}
\caption{IF response of the 230~GHz ({\it a}) and 460~GHz ({\it b}) balanced DSB receivers as measured at representative 
LO frequencies. The standing wave corresponds to the distance between mixer block and isolator \cite{Pamtech} ($\sim$5~cm) .
}
\label{230460IF}
\end{figure*}

To investigate the spurious rejection properties of the 230~GHz and 460~GHz balanced mixers we injected harmonic tones
into the local oscillator path of the respective mixers by means of a second Ka-band synthesizer, connected in parallel 
(via a Ka-band 20~dB directional coupler) to the input of the a Pacific Millimeter Products tripler. \footnote{
Pacific Millimeter Products, Inc., ``Models: E3, E3$^{+}$, W3, W3$^{-}$", 64 Lookout Mountain Circuit, Golden, Co 80401, USA.
[Online] Available: http://www.pacificmillimeter.com/} \cite{kooi_2012}
The result is shown in Fig. \ref{230460Spurs}. Everything being the same, the mixers were biased either symmetric (in phase)
or antipodal. This experiment was done at a number of LO settings. On average the spurious rejection
measured in this manner was 10.1~$\pm$ 2.8~dB for
the 230~GHz balanced mixer and 11.9~$\pm$ 1.5~dB for the 460~GHz balanced mixer. This method proved to be tedious however, 
so we repeated the measurement in a manner described by Westig {\it et. al} \cite{Westig_2011, Westig_2012}.

In a balanced mixer with the IF signals connected to a 180$^\circ$ hybrid coupler, the down-converted RF and LO noise signals 
end up either at the summing node ($\Sigma$) or difference node ($\Delta$), depending on the bias scheme.
Referring to Fig. \ref{Balmixer}, the CSO balanced receivers with integrated IF summing node do not have direct
access to a difference node. However by switching the bias from antipodal to symmetric either the
down-converted RF, or LO noise signal ends up at the mixer ($\Sigma$) output port, as shown in Fig. \ref{NR}a, b. It should be noted that
the last LO multiplier and fixed tuned attenuator are mounted on the cryostat LHe work surface and thermally strapped to the
15~K stage (Fig \ref{incryostat}b). Thus the thermal noise contribution of the LO multiplier/attenuator may be assumed negligible
compared to amplitude noise present from external sources on the LO carrier. 
The balanced mixer noise rejection may be obtained as follows

\begin{equation}
\label{NRmeasured}
NR_{meas}(dB) = -10 \cdot log \frac{(Phot-Pcold)^{+}}{(Phot-Pcold)^{-}}~.
\end{equation}

\noindent
Here '+' indicates symmetric bias and '-' antipodal bias. Phot and Pcold are the averaged receiver IF output signals over 4~GHz
of output bandwidth. The modulus of the antipodal bias equals the symmetric bias, and corresponds to 
that of the obtained receiver sensitivity of Fig. \ref{230Trec}, \ref{460Trec}.
The measurement was done in automated fashion in 100 steps across the 230~GHz and 460~GHz frequency bands, with the 
derived noise rejection plotted in Fig. \ref{NR}c, d.

From the discussion and derived results it is evident that the balanced receiver is capable of suppressing a significant amount of close in amplitude noise
and spurious content (see also section \ref{sect:stab}). A YIG tracking filter \cite{kooi_2012} (or equivalent) is still needed however to remove
AM and spurious noise far from the LO carrier, which can have very large amplitude.

\subsection{IF Response}
\label{sect:IFresponse}

As part of a spurious investigation, the (DSB) IF response of the
230~GHz \& 460~GHz receivers was obtained by stepping the LO frequency between 180$-$280~GHz 
and 390$-$520~GHz in 4~GHz steps. The data was obtained with a FFTS  at an resolution of 256~MHz. 
A typical 230~GHz \& 464~GHz receiver IF spectrum
is shown in Fig. \ref{230460IF}. As discussed in section \ref{sect:IntegratedIFmatching}, the twin-junction SIS design
affords a 1$-$13~GHz IF passband response. However for practical reasons
we have opted for a 4 GHz passband, though in principle an upgrade is possible.

\subsection{Instrument Stability} 
\label{sect:stab}  

\begin{figure*}[t]
\centerline{
\includegraphics[width=110mm]{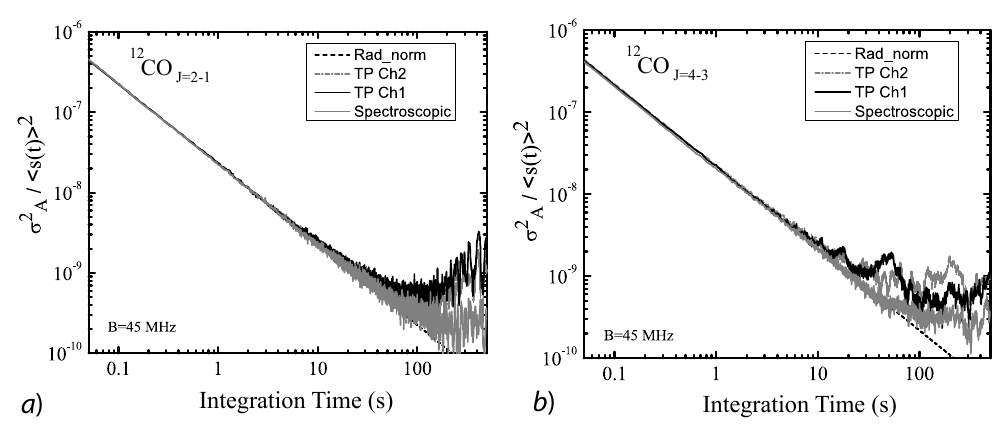}
}
\caption{Measured 230~GHz ({\it a}) and 460~GHz ({\it b}) total-power (continuum) and
spectroscopic Allan variance as a function of
integration time with the instrument mounted on the telescope and looking into an ambient temperature load. The measurement system
noise fluctuation bandwidth was 45~MHz. The 230/460~GHz total-power and spectroscopic Allan variance times ($T_A$) are
approximately 80s/200s and 50s/150s respectively. 
}
\label{AllanVar}
\end{figure*}

Instrument stability, by means of amplitude noise and spurious tone mitigation, is one of the driving motivations 
for the use of balanced receivers. It is found for example that 
poor receiver stability leads to a loss in integration efficiency, poor baseline quality \cite{Kooi_Allan},
and negatively effects observation modes such as ``on-the-fly" mapping and ``drift-scans" as more ``off-source" observations are required.

Throughout the balanced receiver design process \cite{kooi_2012}, much attention has been given to the multiplicity 
of factors that degrade the instrument stability. These include improved SIS and LNA
bias electronics, voltage-divider networks in the SIS mixer and cryogenic
low-noise amplifiers, enhanced thermal design of the room-temperature IF amplifiers,
careful elimination of all ground loops, the use of twisted-pair wires in the cryostat to minimize 
electromagnetic interference (EMI),
and the physical mounting of LNAs and last stage multiplier in a low vibration environment.

The resulting Allan variance stability plot is shown in Fig. \ref{AllanVar}. It has been found \cite{Schieder, Volker} 
that fluctuations with a $f^{-\alpha}$ power spectrum show up in the Allan variance plot as $T_{int}^{\alpha-1}$, 
with $T_{int}$ defined as the  integration time. If we let $\beta = \alpha -1$, the 
shape of the Allan variance is found to follow 

\begin{equation}
\label{AllanShape}
\sigma_A^2 (T_{int}) = a T_{int}^{-1} +  b + c T_{int}^{\beta} ~,
\end{equation}

\noindent
where a, b, c are constants. The first term, with $\beta$~=~-1, represents radiometric (white) noise. In a 
log-log plot it has a slope of -1 (Fig. \ref{AllanVar}).
This type of frequency-independent (uncorrelated) noise integrates down with the square-root of time according to the 
well-known radiometer equation \cite{Krauss} 

\begin{equation}
\label{radiometer}
\sigma =\frac{<s(t)>}{\sqrt{B T_{int}}}~.
\end{equation}

\noindent
s(t) is the measured detector IF output signal in the time domain, and B the equivalent IF noise-fluctuation
bandwidth of the system. The last term in Eq. \ref{AllanShape} represents drift noise with drift index $\beta$.
In between these two limits a certain amount of gain-fluctuation or flicker noise with a 1/f noise power spectral distribution
exists. 
The Allan minimum time  ($T_A$)  may be defined as the intercept between radiometric and drift noise \cite{Schieder}.
However even without gain fluctuation (1/f) noise the minimum in the Allan variance plot is already significantly above the noise level predicted by the radiometer equation. Frequently however significant amounts of LO induced 1/f  noise is present at the radiometer output and a more useful definition of the 
Allan minimum time is the integration time for which the measured noise exceeds that predicted from the radiometer equation by a factor $\sqrt{2}$ \cite{Volker}.

If the stability were to be limited by drift noise alone, the Allan variance time scales with bandwidth \cite{Schieder} as

\begin{equation}
\label{AllanBW}
T_{A}' = T_A (B/B')^{\frac{1}{1+\beta}}~.
\end{equation}

\noindent
Referring the results of Fig.~\ref{AllanVar} to a 1~MHz spectrometer channel bandwidth, the 230~GHz and 460~GHz 
total-power and spectroscopic Allan variance times would be $\sim$500s/1350s and $\sim$335s/1000s respectively.
This is with a fully synthesized LO \cite{kooi_2012}. The measured results of Fig.~\ref{AllanVar} also show that LO noise
with a typical 1/f spectral distribution ($\beta$=0) is not present, supporting the argument that balanced 
receiver noise immunity aids instrument stability. Note that electronic drift noise is present on longer time scales, 
as ordinarily would be the case.

\begin{figure}[t!]
\centerline{
\includegraphics[width=87mm]{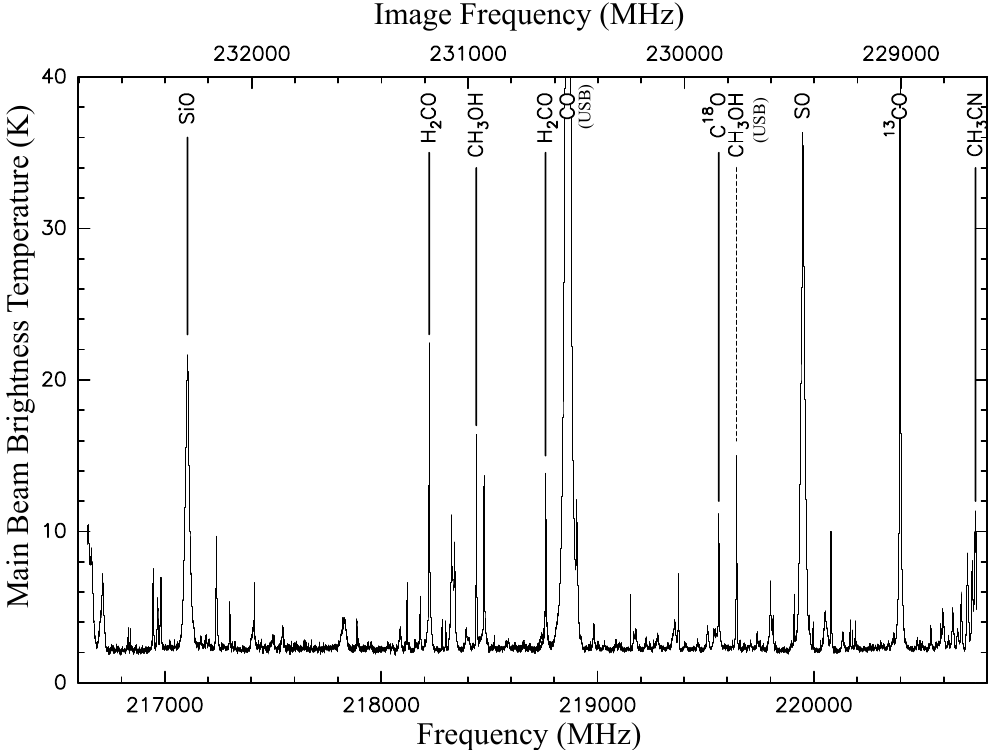}
}
\caption{Spectrum of Orion KL at 219~GHz ($\nu_{LO}$=224.7~GHz). Strongest lines from the signal 
(lower) sideband and image (upper) sideband are identified. 
The Orion spectrum is plotted at the full FFTS resolution of 0.27~MHz. 
The rms noise is $\sim$65~mK rms.
}
\label{orion230}
\end{figure}

For comparison, the ALMA \cite{Alma490} specified goal for total-power gain stability ({\small $\partial$G/G}) at 1s 
is 10$^{-4}$ (B=4~GHz). The results presented here equate to a normalized total-power gain stability ($\sigma/<s(t)>$) 
of $\sim$6.4~x~10$^{-6}$ or roughly 18$\times$ below the ALMA specification.

\subsection{Observations}

In February 2013 we observed Orion KL, the closest high-mass star-forming region. Fig. \ref{orion230} shows a spectrum of Orion KL
between 216.6 and 220.8~GHz. Due to its proximity ($\sim$437~pc) this source exhibits an extremely rich
molecular spectrum across the whole submillimeter band.
The frequency range shown includes the J=2$�-$1 rotational transition of carbon monoxide, CO, together with its two rare
isotopologues, $^{13}$CO and C$^{18}$O. In addition, lines of simple diatomic molecules silicon and sulfur monoxide, SiO and SO, 
are seen, as well as those of formaldehyde, H$_{2}$CO, methanol, CH$_{3}$OH, and many other more complex species. 

The Orion KL spectrum was taken at 52$^{\circ}$ elevation, with a respective single-sideband (SSB) system temperature
and total on-source integration time of 215~K and 2.7~min.
Measured beam efficiency was 73\%, consistent with a 11.5~dB Gaussian illumination (Section \ref{sect:optics}).
The 225~GHz zenith atmospheric opacity was $\sim$0.06, which translates into upper and lower sideband, 
air-mass corrected, on-source opacities of 0.075, 0.073 \cite{Pardo}.

\begin{figure}[t!]
\centerline{
\includegraphics[width=87mm]{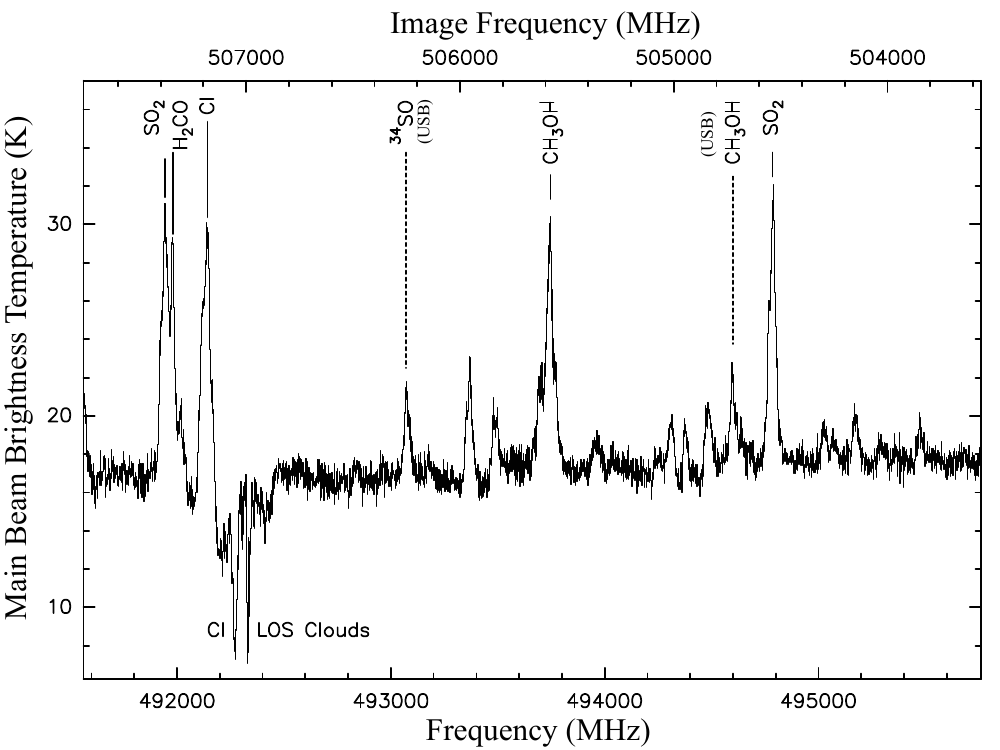}
}
\caption{Spectrum of Sagittarius B2(M) at 493~GHz ($\nu_{LO}$=499.7~GHz). Strongest lines from the (lower) 
sideband and image (upper) sideband are identified. A complex pattern of [C{\footnotesize I}] absorption in 
the foreground clouds on the line of sight between the Sun and the Galactic center can be seen.
The Sgr B2 spectrum is smoothed to 1.1~MHz to lower the noise ($\sim$400~mK rms).
}
\label{Seg492}
\end{figure}

The measured SSB system temperatures (T$_{sys}^{SSB}$) are consistent with those obtained from theory \cite{Jewell}

\begin{equation}
\label{TsysDSB}
T_{sys, SSB}^{DSB, lsb}(\nu) = \frac{\bigl[2T_{rec}^{DSB}(\nu)+T_{ant}^{usb}(\nu)+T_{ant}^{lsb}(\nu)\bigr]}{\eta_s \cdot e^{-\tau^{lsb}(\nu)}}~.
\end{equation}

\noindent
With the antenna brightness temperature 

\begin{equation}
\label{Tant}
T_{ant}(\nu) = \eta_s T_{sky} \bigl[1-e^{-\tau(\nu)}\bigr] + (1-\eta_s)T_{spill} + \eta_sT_{cbg}e^{-\tau(\nu)}~.
\end{equation}

\noindent
$\tau$ is the on-source upper or lower sideband mean opacity. The physical temperature 
of the sky (T$_{sky}$) and antenna spillover temperature ($T_{spill}$) are
estimated to be $\sim$275~K. $T_{cbg}$ is the cosmic background 
temperature (2.726~K), and $\eta_s$ the antenna hot spillover efficiency ($\sim$90\%).
Given a 40~K DSB receiver noise temperature (Fig.~\ref{230Trec}) we obtain a theoretical
T$_{sys}^{SSB}$ of 208~K. In the event a SSB receiver with 10~dB sideband rejection (ALMA) and 
T$_{rec}^{SSB}$~=~2T$_{rec}^{DSB}$ had been used for the observations the SSB system temperature is estimated
to have been 158~K. The gain in system noise temperature by using a SSB receiver 
would have been higher in ``bad" weather observation conditions. 

Fig.~\ref{Seg492} shows a spectrum of Sagittarius B2(M) between 491.8 and 495.8~GHz taken in May 2012.
Sagittarius B2 is the most massive 
molecular cloud complex in the Galactic center and an active region of high-mass star formation. Lines of sulfur monoxide,
$^{34}$SO, sulfur dioxide, SO$_{2}$, formaldehyde, and methanol, can be identified in the spectrum. 
In addition the atomic fine structure line of neutral atomic carbon, [C{\footnotesize I}], is seen in emission at velocities 
corresponding to the envelope of Sagittarius B2. A complex pattern of [C{\footnotesize I}] absorption is also seen at 
velocities corresponding to the foreground molecular clouds located in the Milky Way spiral arms
between the Sun and the Galactic center. 

The Sagittarius B2 spectrum was taken at 42$^{\circ}$ elevation, with a respective SSB system temperature
and total on-source integration time of $\sim$3900~K and 11~min.
Measured beam efficiency was 40\%, consistent with a 11.5~dB telescope illumination (Section \ref{sect:optics})
and 24~$\mu$m surface rms (the CSO surface correction system DSOS was not in use). Weather conditions were marginal for this wavelength band, 
$\tau_{225}$ being $\sim$0.06. These conditions translate
into (averaged) upper and lower sideband, air-mass corrected, on-source opacities of 1.82 and 2.4 \cite{Pardo}. 
If a 45~K DSB receiver noise temperature is assumed (Fig. \ref{460Trec}) we obtain a theoretical
T$_{sys}^{SSB}$ of 3950~K. Again in the event a SSB receiver with 10~dB sideband rejection (ALMA) and 
T$_{rec}^{SSB}$~=~2T$_{rec}^{DSB}$ had been used for the observations, the SSB system temperature is estimated
to have been 2400~K. This is a significant improvement (a factor 2.7 in integration time), 
arguing for the development of balanced SSB mixers. In the case of DSB receivers the IF should be chosen
to minimize the atmospheric noise contribution from the image sideband.

\section{Conclusion}
To facilitate deep integrations, automated line surveys, and remote observations a suite of fully synthesized dual-color
balanced receivers covering the 180$-$720 GHz submillimeter atmospheric frequency range (ALMA B5$-$B9) have
been developed for the Caltech submillimeter Observatory. It was judged an optimal compromise 
between scientific merit and finite funding. In the development wide RF bandwidth was favored \cite{kooi_2007, kooi_2012}, 
allowing the same science to be done with fewer instruments. 


High-current-density (25~kA/cm$^2$) AlN-barrier SIS technology has been used, facilitating 
the very wide instantaneous RF bandwidth presented. For the 230~GHz balanced receiver we obtain in the 180$-$280~GHz frequency range
a receiver sensitivity of 33$-$50~K DSB. The mixer conversion gain is relatively constant at 0~$\pm$~1 dB. The 
460~GHz balanced receiver also has a noise temperature in the range 40$-$50~K DSB, with a mixer conversion gain of 1~$\pm$~1 dB. 
In the case of the 460~GHz receiver the very low receiver noise temperature is understood to be the result if four factors: A cooled last
stage multiplier, $\sim$12~dB of AM noise rejection, moderate mixer conversion gain, and high optical throughput. 
Conversely, the 230 GHz receiver noise temperature suffers from the fast optics required to achieve proper
telescope illumination and less of a reduction in LO noise as is the case with the 460~GHz instrument.

The amplitude noise rejection is 12~dB $\pm$~3~dB for both receivers. This result 
slightly exceeds the theoretical estimate of \cite{kooi_2012}. Instrument stability, 
as measured in situ at the telescope, is excellent and argues for the development of submillimeter 
and terahertz balanced receiver configurations. The presented results are confirmed by actual 
observations and the obtained high quality spectral baselines. 

Unfortunately, deployment of the 345/650 balanced receiver(s) is presently on hold due to funding difficulties.

\section{Acknowledgements}
The authors wish to thank J. Groseth and D. Warden, California Institute of Technology, for the assembly of the 
needed bias electronics. Prof. S. Weinreb, Jet Propulsion Laboratory and California Institute of Technology, 
for making available the cryogenic low noise MMIC's, Dr. J. Pierson of the Jet Propulsion Laboratory for
his assistance with the medium power amplifiers modules, Prof. P. Goldsmith of the Jet Propulsion Laboratory for his advice
on instrument stability and general support, and Prof. J. Zmuidzinas of the California Institute of Technology
for providing the K$_a$-band synthesizers, the wideband Fast Fourier Transform Spectrometers (FFTS),
and for his advise and physics insight over the years.

\nocite{*}
\bibliographystyle{IEEE}

%
\begin{IEEEbiography}[{\includegraphics[width=1in,height=1.25in,clip,keepaspectratio]{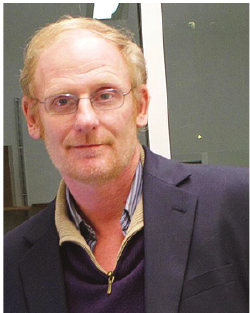}}]{Jacob W. Kooi}
was born in Geldrop, The Netherlands on July 12, 1960. He received his B.S. degree in Microwave Engineering at the 
California Polytechnic State University in San Luis Obispo, California in 1985, a M.S. degree in Electrical Engineering from the 
California Institute of Technology in 1992, and Ph.D in Physics at the University of Groningen, the Netherlands in 2008.
His research interests are in the area of Millimeter and Submillimeter wave technology, low energy physics, electrodynamics, thermodynamics, low noise amplifiers and associated MMIC technology,
Fourier optics, instrumental stability, and their application to astronomy and aeronomy. He is currently with the Combined Array 
for Research in Millimeter-wave Astronomy (CARMA) at the California Institute of Technology.
\end{IEEEbiography}

\begin{IEEEbiography}[{\includegraphics[width=1in,height=1.25in,clip,keepaspectratio]{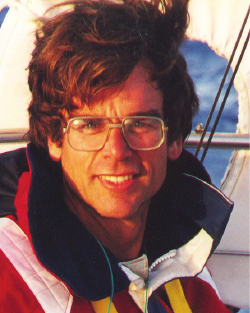}}]{Richard Chamberlin} 
was graduated from the University of California (Santa Barbara, CA) with a B.S. in physics in 1984, and obtained his Ph.D. in
physics from the Massachusetts Institute of Technology (Cambridge, MA) in 1991 under George B. Benedek. He served in the United States Air 
Force from 1975 to 1979 as a Weather Observer. In 1995 he was the first winter-over scientist with the
pioneering Antarctic Submillimeter Telescope and Remote Observatory which he helped design, build, and test while at Boston University. From 1996 to
2010 he was the Technical Manager of the Caltech Submillimeter Observatory. His research interests include water vapor in the atmosphere and terahertz remote sensing. 
He is currently affiliated with the NIST lab in Boulder, CO.
\end{IEEEbiography}

\begin{IEEEbiography}[{\includegraphics[width=1in,height=1.25in,clip,keepaspectratio]{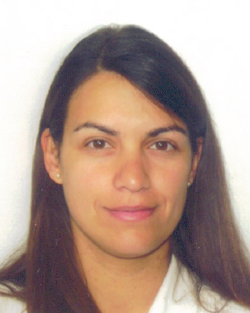}}]{Raquel R. Monje}
received the M.S. degree in telecommunication engineering from Universidad Europea de Madrid, Spain in 2003, the M.S. 
degree in digital communications system and technology from Chalmers University of Technology, Gothenburg, Sweden in 2004, and the Ph. D. 
degree in radio and space science from Chalmers University of Technology, Gothenburg, Sweden in 2008. Her Ph.D thesis was on low noise 
superconductor-insulator-superconductor (SIS) mixers for submillimeter and millimeter-wave astronomy.

She is currently a senior postdoctoral scholar at California Institute of Technology, Pasadena, CA. Her research interests include microwave 
technology, SIS mixers, millimeter and submillimeter wave heterodyne receivers for astronomy and the associated science resulting from observations.
\end{IEEEbiography}

\begin{IEEEbiography}[{\includegraphics[width=1in,height=1.25in,clip,keepaspectratio]{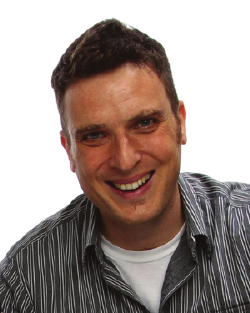}}]{Attila Kov\'{a}cs}
graduated from Harvard University with an A.B. in physics, astrophysics, and astronomy in
1997, then received his Ph.D. in physics from Caltech in 2006, in the submillimeter astrophysics group
lead by Tom G. Phillips. As a student, he pioneered the observing modes (e.g. Lissajous scans) and the
most widely used data reduction approach (CRUSH) for ground-based total-power imaging arrays in the
submillimeter, and designed the SIS mixers for this work. Later, he worked at the Max Planck Institute
for Radioastronomy in Bonn as post-doc, under the mentorship of Karl Menten, taking in a lead in the
optimization and commissioning of the APEX bolometer cameras LABOCA (and its polarimetry frontend PolKa)
and SABOCA and conducting surveys with them. In 2009 he became an independent postdoc at the University
of Minnesota, focusing on improving and commissioning the GISMO 2-mm camera for the IRAM 30-m telescope,
and providing the concept of lithograhpic on-chip spectrometers for the (sub)millimeter. Since 2012 he
has been back at Caltech as a post-doc with Jonas Zmuidzinas, working on SuperSpec and the MAKO KID
camera. He is also part of the ongoing collaborations for the SOFIA HAWC+ upgrade and the GISMO-2 camera.
Beyond the technical work, he is also very interested in studying the high-z mm-bright star-forming
populations and empirical dust models for understanding them.
\end{IEEEbiography}

\begin{IEEEbiography}[{\includegraphics[width=1in,height=1.25in,clip,keepaspectratio]{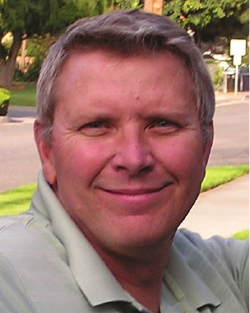}}]{Frank Rice }
has been Lecturer in Physics at California Institute of
Technology since 2001. He received his physics B.A. from Northwestern
University in 1977 and earned his M.S. in physics from Caltech in 1997.
Prior to coming to Caltech he served as a light-attack aviator and test
pilot in the U.S. Navy. He graduated from Naval Postgraduate School and the
U.S. Naval Test Pilot School with a M.S. in aerospace engineering in 1986;
he later served as a flight test instructor pilot at USNTPS. At Caltech he
has specialized in superconducting physics for millimeter and submillimeter
heterodyne receiver design and designed a wide bandwidth, fixed-tuned SIS
receiver for 230 GHz which was used as a CSO facility instrument for several
years. He designed the operating and bias electronics for this new suite of
CSO heterodyne instruments.
\end{IEEEbiography}

\begin{IEEEbiographynophoto}{Hiroshige Yoshida} received his M.S. degree in Physics from The
University of Tokyo, Tokyo, Japan in 1994. In 1997, he became a staff
member of the Caltech Submillimeter Observatory in Hawaii, and has
since been developing and maintaining observatory's software systems.
\end{IEEEbiographynophoto}

\begin{IEEEbiographynophoto}{Brian Force} 
is a RF/microwave engineer. Previous technical duties included work at the Caltech Submillimeter Observatory (CSO) 
and James Clerk Maxwell Telescope (JCMT), both located on Mauna Kea, Hawaii. Part of his assignment included 
support of the telescopes and help observers operate it. Currently Brian is working for a small high tech company in California designing synthesizers and RF systems.
\end{IEEEbiographynophoto}

\begin{IEEEbiography}[{\includegraphics[width=1in,height=1.25in,clip,keepaspectratio]{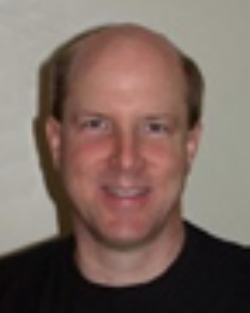}}]{Kevin Cooper}
is an electronics and software engineer who works at the California Institute of Technology.
Part of his duties included designing hardware and software for testing and characterization of 
the balanced SIS receivers for the Caltech Submillimeter Observatory.
\end{IEEEbiography}

\begin{IEEEbiography}[{\includegraphics[width=1in,height=1.25in,clip,keepaspectratio]{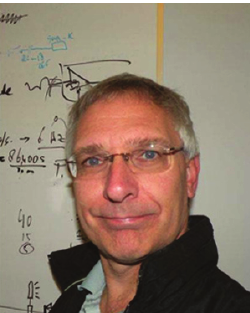}}]{David Miller}
David Miller received his B.S. and M.S. in Electrical Engineering at California State
Polytechnic University, Pomona, in 1990 and 1998, respectively. His interests include microwave engineering,
low noise and high-stability electronics, and the design, construction, and testing of submillimeter receivers
for airborne, satellite, and land-based observatories. He has recently graduated from Talbot School of Theology
with M.A. in Theology. Currently he is working for Nuvotronics, LLC, where he designs, builds, tests, and
characterizes next-generation multi-watt solid-state power amplifiers in the microwave and millimeter-wave
frequency bands. 
\end{IEEEbiography}

\begin{IEEEbiography}[{\includegraphics[width=1in,height=1.25in,clip,keepaspectratio]{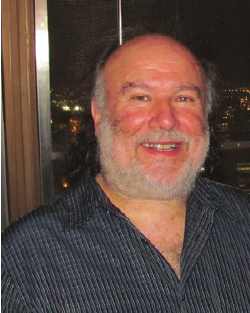}}]{Marty Gould}
spent 12 years at California Institute of Technology as an Instrument
Specialist Machinist, and taught machine shop in the Mechanical Engineering
Department.  He started Zen Machine and Scientific Instrument in 1985 and moved to
Colorado in 1992 where he specializes in scientific research support along with
occasionally mentoring high school students in machine shop. 
\end{IEEEbiography}

\begin{IEEEbiography}[{\includegraphics[width=1in,height=1.25in,clip,keepaspectratio]{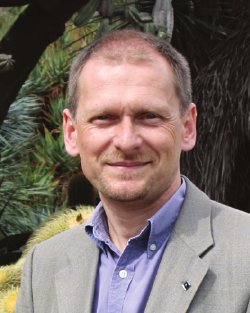}}]{Dariusz Lis}
received his Ph.D. from the University of Massachusetts at Amherst in 1989. He is Senior Research Associate in Physics at the
California Institute of Technology and Deputy Director of the Caltech Submillimeter Observatory. His research interests include volatile
composition of comets, astrochemistry, molecular spectroscopy of the interstellar medium, isotopic fractionation and deuteration,
photon-dominated regions, far-IR continuum and spectroscopic studies of star-forming regions, as well as the ISM in the high-z universe. 
\end{IEEEbiography}

\begin{IEEEbiography}[{\includegraphics[width=1in,height=1.25in,clip,keepaspectratio]{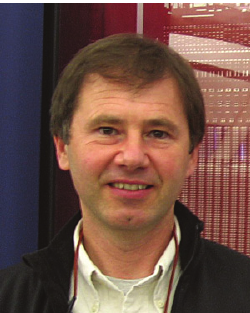}}]{Bruce Bumble}
received a B.E. degree in Engineering Physics from Stevens Institute of Technology (Hoboken, NJ) in 1982 and a M.S degree in Material Science 
from Polytechnic University (Brooklyn, NY) in 1989.  He worked on superconducting device fabrication applications in computing at IBM Watson 
Research and from 1989 to the present is a Member of Technical staff at the Jet Propulsion Laboratory in Pasadena, CA working on
superconducting materials and devices mainly for astronomy applications.  This includes fabricating SIS and Hot Electron Bolometers for
heterodyne mixers from Nb, NbN and NbTiN materials.  Current work includes MKID arrays for optical cameras and TES Bolometers for mm wave spectrometers.
\end{IEEEbiography}



\begin{IEEEbiography}[{\includegraphics[width=1in,height=1.25in,clip,keepaspectratio]{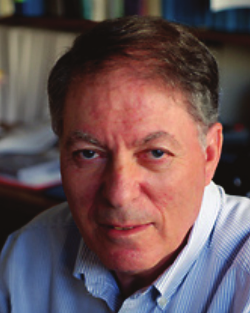}}]{Tom G. Phillips}
was educated at Oxford, England, where he received B.A., M.A., and D.Phil. degrees. His graduate studies were in 
low-temperature physics. After one year at Stanford University, he returned to Oxford for two years and then moved 
to the Bell Laboratories Physics Research Laboratory at Murray Hill, NJ.
There he developed techniques for millimeter and submillimeter wave detection for astronomy. In 1975 he 
spent one year at London University as University Reader in Physics. 
In 1980 he joined the faculty of Caltech as Professor of Physics. 
At Caltech he took on the task of construction of the Owens Valley Radio Observatory millimeter wave interferometer, as 
Associate Director of the Observatory. In 1982 he became Director Designate for the Caltech Submillimeter wave Observatory, 
to be constructed in Hawaii, and in 1986, on successful completion of the construction,
became Director. His current research interests are in molecular and atomic spectroscopy
of the interstellar medium and in the development of superconducting devices for submillimeter-wave detection.
\end{IEEEbiography}

\end{document}